\documentclass[sigconf,natbib,screen=true,review=false,anonymous=false]{acmart}

\AtBeginDocument{%
  \providecommand\BibTeX{{%
    \normalfont B\kern-0.5em{\scshape i\kern-0.25em b}\kern-0.8em\TeX}}}

 \copyrightyear{2023} 
 \acmYear{2023} 
 \setcopyright{acmlicensed}\acmConference[KDD '23]{Proceedings of the 29th ACM SIGKDD Conference on Knowledge Discovery and Data Mining}{August 6--10, 2023}{Long Beach, CA, USA}
 \acmBooktitle{Proceedings of the 29th ACM SIGKDD Conference on Knowledge Discovery and Data Mining (KDD '23), August 6--10, 2023, Long Beach, CA, USA}
 \acmPrice{15.00}
 \acmDOI{10.1145/3580305.3599487}
 \acmISBN{979-8-4007-0103-0/23/08}


\usepackage{bm}
\usepackage{makecell}
\usepackage{subfigure}
\usepackage{multirow}

\usepackage{tikz}
\usepackage{subfigure}
\usepackage{mathtools}

\usepackage{subfigure}
\usepackage[most]{tcolorbox}
\usepackage{booktabs}
\usepackage{arydshln}
\usepackage{multirow}
\usepackage{bm}
\usepackage{color}
\usepackage{float}
\usepackage{amsfonts}
\usepackage{mathbbol}
\usepackage{bm}
\usepackage[ruled,vlined,linesnumbered,noresetcount]{algorithm2e}
\usepackage{enumitem}
\usepackage[switch]{lineno}

\theoremstyle{plain}

\newtheorem{theorem}{Theorem}[section]

\newtheorem{lemma}[theorem]{Lemma}

\newcolumntype{I}{!{\vrule width 1pt}}

\newcommand*{\Cline}[1]{%
\noalign{\global\setlength{\arrayrulewidth}{1pt}}%
\cline{#1}%
\noalign{\global\setlength{\arrayrulewidth}{0.4pt}}%
}
\usepackage{xurl}

\usepackage{pifont}
\definecolor{darkgreen}{RGB}{208,64,56}

\newcommand{\cmark}{{\color{darkgreen}\ding{51}}}
\newcommand{\xmark}{\ding{55}}

\begin{document}

\title{Reconsidering Learning Objectives in Unbiased Recommendation with Unobserved Confounders}


\author{%
Teng Xiao
}
\affiliation{%
  \institution{
   The Pennsylvania State University
  }
   \country{}
}  
\email{tengxiao@psu.edu}

\author{%
Zhengyu Chen
}
\affiliation{%
  \institution{
   	Zhejiang University
  }
   \country{}
}  
\email{chenzhengyu@zju.edu.cn}

\author{%
Suhang Wang
}
\affiliation{%
  \institution{
  The Pennsylvania State University
  }
   \country{}
}  
\email{szw494@psu.edu}

\begin{abstract}
This work studies the problem of learning unbiased algorithms from biased feedback for recommendation. We address this problem from a novel distribution shift perspective. Recent works in unbiased recommendation have advanced the state-of-the-art with various techniques such as re-weighting, multi-task learning, and meta-learning. Despite their empirical successes, most of them lack theoretical guarantees, forming non-negligible gaps between theories and recent algorithms. In this paper, we propose a theoretical understanding of why existing unbiased learning objectives work for unbiased recommendation. We establish a close connection between unbiased recommendation and distribution shift, which shows that existing unbiased learning objectives implicitly align biased training and unbiased test distributions. Built upon this connection, we develop two generalization bounds for existing unbiased learning methods and analyze their learning behavior.
Besides, as a result of the distribution shift, we further propose a principled framework, Adversarial Self-Training (AST), for unbiased recommendation. Extensive experiments on real-world and semi-synthetic datasets demonstrate the effectiveness of AST.
\end{abstract}

\maketitle

\section{Introduction}
\label{sec:intro}
Recommender systems are widely used in many applications such as e-commerce platforms, social networks, and healthcare. However, recommender systems learn from logged user-item feedback data and are subject to selection bias as the training data collected by the logging policy is observational rather than experimental~\cite{yang2018unbiased, schnabel2016recommendations,xiao2021general, DBLP:conf/icml/GuptaWLW21}. Ideally, the feedback should be collected by randomly and uniformly exposing items to users. However, in the real world, exposures are affected by the past recommendation policy, which is known as model selection bias. For example, users are more likely to interact with popular items than tail items, and recommender systems are also more likely to recommend popular items than others~\cite{yang2018unbiased, schnabel2016recommendations,xiao2022towards}. This model selection bias results in the "rich get richer" phenomenon, where head contents are getting more and more exposure while tail contents are rarely discovered. Selection bias also comes from user self-selection, i.e., users usually interact and rate items they like and rarely rate items they do not like~\cite{marlin2007collaborative, schnabel2016recommendations}. Previous studies~\cite{schnabel2016recommendations,saito2020unbiased, wang2019doubly,xiao2022representation} have theoretically and empirically shown that directly learning from the biased feedback cannot reflect user true preferences on items.

Remarkable theoretical advances have been proposed for unbiased recommendation. Specifically, \cite{schnabel2016recommendations} and \cite{wang2019doubly} provide rigorous generalization bounds under selection bias. On par with their theoretical findings, there have been rich advances in unbiased recommendation~\cite{schnabel2016recommendations, wang2019doubly, saito2020unbiased, yang2018unbiased} based on inverse propensity score (IPS)\cite{rosenbaum1983central} and doubly robust (DR)\cite{bang2005doubly} in causal inference. Although IPS and DR can address the selection bias in theory, these solutions typically assume unconfoundedness~\cite{xu2020adversarial}, i.e., the independence of user preference over items given the feature of getting exposed~\cite{saito2020unbiased, schnabel2016recommendations, xu2020adversarial}, which is impractical and cannot be examined in many real-world RS. Moreover, they need to estimate the propensity score for re-weighting and suffer from huge variance when the propensity score is small~\cite{swaminathan2015self, saito2020asymmetric}. Thus, IPS and DR empirically perform poorly compared to many recent works~\cite{liu2020general, DBLP:conf/sigir/ChenDQ0XCLY21, wang2021combating, wang2020information, saito2020asymmetric, wan2022cross}.

\begin{table*}[h]
\caption{An overview of representative  unbiased learning objectives we theoretically discuss in this paper, and how they relate to one another in terms of unconfoundedness assumption and whether they can work w/o unbiased uniform data, suffer from the variance issue, or whether the methods can theoretically unify other algorithms.}
\vskip -1.3em
\centering
\setlength{\tabcolsep}{1.5mm}{
\resizebox{0.99\textwidth}{!}{
\begin{tabular}{lcccccccl}
   \toprule[1.0pt]
\textbf{Learning objectives}    &  \textbf{w/o unconfoundedness assumption}    & \textbf{w/o unbiased uniform data}  & \textbf{w/o variance issue} & \textbf{unified framework}  \\
\hline
Re-weighting \cite{schnabel2016recommendations,saito2020unbiased,wang2019doubly,DBLP:conf/sigir/GuoZLYCWCY021}  & \xmark    &\cmark    & \xmark    & \xmark    \\
Information bottleneck \cite{wang2020information,liu2021mitigating} &  \xmark    & \cmark  &  \cmark   & \xmark \\
Multi-task learning \cite{bonner2018causal,liu2020general}  & \xmark    & \xmark &  \cmark    & \xmark  \\
Meta-learning \cite{DBLP:conf/sigir/ChenDQ0XCLY21,wang2021combating}   &  \xmark    & \xmark  &  \cmark  & \xmark 
\\
\hline
Adversarial self-training  & \cmark    & \cmark&  \cmark   & \cmark  \\
\toprule[1.0pt]
\end{tabular}}}\label{overview}
\vskip -1.5em
\end{table*}

Many unbiased recommendation algorithms have been introduced to conduct debiasing learning using various machine learning techniques, such as multi-task learning~\cite{bonner2018causal, liu2020general}, meta-learning~\cite{DBLP:conf/sigir/ChenDQ0XCLY21, wang2021combating}, and information bottleneck~\cite{wang2020information}, which achieve promising empirical performance. However, there is a severe lack of rigorous theoretical analysis for these algorithms in the literature, creating a gap between current theory and many strong empirical methods. Specifically, most of these methods~\cite{liu2020general, wang2021combating, DBLP:conf/sigir/ChenDQ0XCLY21, bonner2018causal} solve the bias issue by introducing unbiased uniform data in the training, which is collected by a random logging policy. Nevertheless, no clear and unified connection between current theory and these algorithms has been established. In other words, unbiased learning generalization bounds for them have not been derived. Furthermore, there is no solid theoretical justification for why utilizing unbiased uniform data can improve learning performance. Table~\ref{overview} provides an overview of the discussed methods and suggests that most of them lack theoretical guarantees. This significant gap between theory and practice raises an important question:
\textit{How to bridge the gap between theories and recent unbiased learning objectives? Furthermore, could we propose a more effective unbiased learning objective guided by rigorous theoretical justification?}


In this paper, we provide answers to the research question stated above. We first revisit unbiased recommendation from the perspective of distribution shift and then present a theoretical analysis of unbiased learning to provide explicit guidance and explanation for the current algorithm design. Our analysis shows that many unbiased learning objectives essentially optimize different terms in our bound. Unlike existing bounds~\cite{schnabel2016recommendations, wang2019doubly}, our bounds explicitly suggest accounting for the unobserved confounders, which is important since the assumption of unobserved confounders may not hold in the real world (please see details in \S~\ref{sec:assump}). Our theoretical generalization bounds pave the way for us to understand why and how unbiased uniform data improves unbiased learning performance. We further provide insights into our theory analysis and propose a novel unbiased learning algorithm, Adversarial Self-Training (AST), which effectively minimizes the upper bound of the error and reduces the unbiased generalization gap. We evaluate AST on both real-world and semi-synthetic datasets and conduct ablation studies to analyze its behaviors. Extensive experimental results validate the effectiveness of AST. The main contributions of this work can be summarized as follows:
\begin{itemize}[leftmargin=*]
   \item We reconsider unbiased learning objectives proposed recently for recommendation from the perspective of distribution shift and provide a novel theoretical analysis towards explicit guidance and explanations for algorithm design.
    \item We provide important insights that our theoretical generalization bounds allow us to understand why and how unbiased uniform data helps to improve unbiased learning performance.
    \item Inspired by our theoretical analysis, we propose a novel unbiased algorithm, AST, which can maintain rigorous theoretical justification and address limitations of current algorithms. Extensive experiments on both semi-synthetic and real-world datasets also demonstrate the effectiveness of AST.
\end{itemize}





\section{Related Work}
\label{sec:RW}
\subsection{Selection Bias in Recommendation}
Unbiased learning algorithms such as IPS~\cite{schnabel2016recommendations,saito2020unbiased,zhu2020unbiased,saito2020unbiased,xiao2022towards} and DR~\cite{wang2019doubly,wang2021combating} are proposed to theoretically address selection bias. For example, DR combines propensity score estimation and error imputation in a theoretically sophisticated manner. However, these methods heavily rely on accurately estimating the propensity score, which is often impossible to know in the real world. Furthermore, previous works~\cite{swaminathan2015self,dudik2011doubly} have demonstrated that these methods suffer from high variance~\cite{sachdeva2020off}. It is important to note that these causal inference methods typically assume unconfoundedness, where the relevance of user-item pairs is assumed to be independent of exposure given the user and item features~\cite{saito2020unbiased,xu2020adversarial,schnabel2016recommendations}. \citeauthor{xu2020adversarial}~\cite{xu2020adversarial} make similar observations regarding the limitations of the unconfoundedness assumption and highlight the inconsistent issues in supervised learning caused by unknown exposure mechanisms. However, they do not provide a theoretical framework to explain existing unbiased learning methods.

Recently, several empirical algorithms have been proposed to avoid the need for estimating the propensity score, utilizing techniques such as causal embedding~\cite{bonner2018causal}, knowledge distillation~\cite{liu2020general,liu2022kdcrec}, and transfer learning~\cite{lin2021transfer}. These algorithms follow a multi-task learning scheme, where both unbiased uniform data and biased data are used, and the difference between the resulting user-item representations is regularized. Additionally, some algorithms adopt a meta-learning scheme~\cite{DBLP:conf/sigir/ChenDQ0XCLY21,wang2021combating}, where unbiased uniform data is used to supervise the learning of debiasing parameters within a bi-level optimization framework. Despite their promising performance in practice, most of these algorithms require additional unbiased uniform data, which can degrade user experiences, and they lack sufficient theoretical guarantees. As a result, there is currently a disconnect between theory and the existing algorithms. This work primarily focuses on addressing selection bias, with the aim of bridging the gap between theories and algorithms by proposing a theoretically motivated framework for unbiased recommendation.

\subsection{Domain Adaptation and Self-Training}
The unbiased recommendation problem setting can be treated as a special instantiation of out-of-distribution generalization and is related to domain adaptation~\cite{ben2007analysis,ben2010theory,ganin2015unsupervised,long2018conditional}. We discuss the relationships of our problem setting and our model with domain adaptation. The goal of domain adaptation is to train a predictor that performs well on a target domain using only labeled source samples and unlabeled target samples during training. The adversarial feature adaptation methods~\cite{ganin2015unsupervised}, inspired by the theoretical analysis of~\cite{ben2007analysis}, are most similar to ours. Specifically, in~\cite{ganin2015unsupervised}, DANN is proposed to simultaneously minimize source empirical errors and approximate the divergence between source and target domains~\cite{ben2007analysis}. Our approach further develops this idea for unbiased learning in recommendation, but our work differs from domain adaptation in :(1) Our work focuses on the unbiased recommendation scenario where both selection bias and unobserved confounders exist simultaneously, as shown in \S~\ref{sec:assump}, and (2) we derive two novel generalization bounds for both multi-task and meta-learning strategies using unbiased uniform data proposed by recent unbiased recommendation algorithms~\cite{liu2020general,wang2021combating,DBLP:conf/sigir/ChenDQ0XCLY21,bonner2018causal}.

Our work is also related to self-training~\cite{grandvalet2004semi,berthelot2019mixmatch,wei2020theoretical}, which is a popular technique for semi-supervised learning. Self-training assigns pseudo-labels to unlabeled samples by using a classifier's predictions and jointly re-trains the model with pseudo-labeled and labeled samples. Instead of focusing on semi-supervised learning, in this paper, we address the unbiased recommendation problem with the self-training. There are also some works~\cite{yuan2019darec,liu2021leveraging,krishnan2018adversarial,chen2020esam} applying the self-training  for long-tail and cross-domain recommendation. Several previous works also have explored adversarial training to improve fairness~\cite{wu2021fairness}, robustness~\cite{wu2021fight}, and accuracy~\cite{wang2017irgan,he2018adversarial} of recommendation. Different from them, we focus on providing a theoretical analysis of existing unbiased learning objectives and addressing the selection bias issue via adversarial self-training.

\section{Preliminaries}
\label{Pre}
In this section, we introduce basic notations  and formulate the unbiased recommendation from the distribution shift perspective. 

\subsection{Notations and Selection Bias}
Let $\mathbf{x}_{u} \in \mathcal{X}_{\mathcal{U}}$ be the feature vector for user $u \in\{1, \ldots, |\mathcal{U}|\}, \mathbf{x}_{i}\in \mathcal{X}_{\mathcal{I}}$ be the feature vector for item $i \in\{1, \ldots, |\mathcal{I}|\}$. Typically, the feature vectors can be user/item one-hot encoding, profile or embedding. $\mathcal{X}_{\mathcal{U}}$ and $\mathcal{X}_{\mathcal{I}}$ are the feature spaces, respectively. Following previous works~\cite{saito2020unbiased,xu2020adversarial}, we let $O_{u, i} \in\{0,1\}$ be the exposure status, $Y_{u, i}\in \{0,1\}$ be the feedback such as the click, and $R_{ui}\in \{0,1\}$ be the true preference of user $u$ on item $i$. $O_{u, i}=1$ if the  feedback $Y_{u, i}$ is observed and $Y_{u, i}=O_{u, i} \cdot R_{u, i}$ which means that, when item $i$  has been exposed to $u$, the true preference should be equal to the feedback~\cite{saito2020unbiased,schnabel2016recommendations}. Let $\mathcal{D}_{P}=\{\mathbf{x}_{u},\mathbf{x}_{i},Y_{ui}| O_{u, i}=1\}$ be the logged feedback and the number of samples  is $N$. The task of unbiased recommendation is to infer unobserved 
preference $R_{ui}$. Typically, the collected feedback follows a  generative process~\cite{saito2020unbiased,schnabel2016recommendations,xu2020adversarial}:
\begin{linenomath}
\small
\begin{align}
&p(\mathbf{x}_{u},\mathbf{x}_{i},Y_{ui})=p(\mathbf{x}_{u})p(\mathbf{x}_{i}) p(R_{ui}, O_{ui}=1| \mathbf{x}_{u},\mathbf{x}_{i})= \\
&p(\mathbf{x}_{u})p(\mathbf{x}_{i}) p(O_{ui}=1|\mathbf{x}_{u},\mathbf{x}_{i}) p(R_{ui}| O_{ui}=1, \mathbf{x}_{u},\mathbf{x}_{i})  \because Y_{ui}=O_{ui}\cdot R_{ui}\nonumber,
\end{align}
\end{linenomath}
where the  exposure distribution $p(O_{ui}=1|\mathbf{x}_{u},\mathbf{x}_{i})$ makes the observed feedback be missing-not-at-random (MNAR). We will drop  $= 1$ for all $O_{ui}$ in the remainder of the paper for conciseness. The exposure distribution $p(O_{ui}|\mathbf{x}_{u},\mathbf{x}_{i})$ is unknown and
depends on user self-selection or the item exposure process by which past-recommendation policies match users and items.

Since we want to eliminate the influence from the underlying exposure mechanism, ideally, we are interested in learning with the following unbiased risk function where the exposure is missing completely at random (MCAR), i.e., $O_{ui} \perp (R_{ui},\mathbf{x}_{u},\mathbf{x}_{i})$:
\begin{linenomath}
\small
\begin{align}
\mathcal{L}_{Q}(f)\triangleq \mathcal{L}_{Q}(f,g)=\mathbb{E}_{Q}[\ell(f(\mathbf{x}_{u}, \mathbf{x}_{i}), g(\mathbf{x}_u, \mathbf{x}_i))] \label{Eq:unbiased}
\end{align}
\end{linenomath}
where $Q\triangleq p(\mathbf{x}_{u})p(\mathbf{x}_{i})p(O_{ui})$ with $p(O_{ui})=1$ for all user-item pairs~\cite{schnabel2016recommendations,xu2020adversarial,xu2021rethinking}. $f(\mathbf{x}_u, \mathbf{x}_i)$ is the estimated hypothesis.  $g(\mathbf{x}_u, \mathbf{x}_i) = p({R}_{ui}|\mathbf{x}_{u},\mathbf{x}_{i})$ is the optimal labeling function, depending on the true preference distribution $p(R_{ui}|\mathbf{x}_u,\mathbf{x}_i)$. $Q$ is called as the marginal distribution over features. Typically,  $\ell(f(\mathbf{x}_u, \mathbf{x}_i), g(\mathbf{x}_u, \mathbf{x}_i))$ is the  0-1 loss, which is the probability that $f$ disagrees with $g$ under Q: $\mathbb{E}_{Q}[\mathbf{I}(f(\mathbf{x}_u,\mathbf{x}_i) \neq g(\mathbf{x}_u,\mathbf{x}_i))]$. In this paper, we conduct theoretical analysis based on 0-1 loss. But, in practice, we can use 0-1 log loss $\ell(x,y)=-y\log\sigma(x)-(1-y)\log(1-\sigma(x))$ with $\sigma(x)=1/(1+e^{-x})$ which serves as a effective convex proxy for 0-1 loss.

We can notice that the unbiased risk function in Eq.~\eqref{Eq:unbiased} is independent of
the exposure distribution of  logged feedback, i.e., $p(O_{ui}|\mathbf{x}_{u},\mathbf{x}_{i})$. That is, we average the instance-wise loss over the uniform exposure distributions of all user-item pairs, $P(O_{ui})=1$,
rather than the  exposure distribution $p(O_{ui}|\mathbf{x}_{u},\mathbf{x}_{i})$. This uniform exposure scenario is ideal because the preference will not be affected by the previous exposure, thus leading to an unbiased estimation. In other words, unbiased recommendation wants to learn hypothesis $f$ which generalizes well for all possible pairs of users and items, not just the pairs that are frequently exposed.
The reason we suffer from the bias  is because of the discrepancy between the exposure distribution of  the logged feedback, and
the testing distribution to which the model will be practically applied:
\begin{linenomath}
\small
\begin{align}
&\textbf{Training}: {p=p(\mathbf{x}_u)  p(\mathbf{x}_{i}) p(O_{u i}| \mathbf{x}_{u}, \mathbf{x}_{i})} p(R_{u i} | O_{u i}, \mathbf{x}_{u}, \mathbf{x}_{i})\\ &\textbf{Testing}: ~q=p(\mathbf{x}_u) p(\mathbf{x}_{i})p(O_{ui})p(R_{ui}|\mathbf{x}_u,\mathbf{x}_i) ~Y_{u i}=O_{u i} \cdot R_{u i} \label{Eq;shift}
\end{align}
\end{linenomath}
Thus, the empirical risk $\mathcal{\widehat{L}}_{P}(f)$ over  logged feedback $\mathcal{D}_{P}$ is a biased estimate of the ideal risk:
\begin{linenomath}
\small
\begin{align}
&\hat{\mathcal{{L}}}_{P}(f)=\frac{1}{N}  \sum\nolimits_{ (\mathbf{x}_{u},\mathbf{x}_{i},Y_{ui}) \in \mathcal{D}_{p}}  \ell (f(\mathbf{x}_u, \mathbf{x}_i),Y_{ui}) \simeq \mathcal{L}_{P}(f) \neq \mathcal{L}_{Q}(f),  \nonumber \\ 
&\mathrm{where}~ \mathcal{L}_{P}(f)\triangleq  \mathcal{L}_{P}(f,k) =\mathbb{E}_{P}[\ell(f(\mathbf{x}_{u}, \mathbf{x}_{i}), k(\mathbf{x}_u,  \mathbf{x}_i))], \label{Eq:LP}
\end{align}
\end{linenomath}
$P=p(\mathbf{x}_{u})p(\mathbf{x}_{i}) p(O_{ui}=1|\mathbf{x}_{u},\mathbf{x}_{i})$ and $k(\mathbf{x}_u, \mathbf{x}_i)$ is the optimal labeling function depending on  distribution  $p(R_{ui}|\mathbf{x}_{u},\mathbf{x}_{i},O_{ui})$ on training. 
Thus, the learned  $f$ will not be approximately optimal even having sufficiently large training data~\cite{schnabel2016recommendations}.


\subsection{The Unconfoundedness Assumption}
\label{sec:assump}
To deal with this selection bias, many de-biasing methods~\cite{schnabel2016recommendations,saito2020unbiased,wang2019doubly} inspired by causal inference algorithms such as IPS and DR have been proposed. As mentioned by previous works~\cite{saito2020unbiased,xu2020adversarial}, these algorithms assume that being relevant is independent of getting exposed given the feature, i.e, $R_{ui} \perp O_{ui}| \mathbf{x}_{u},\mathbf{x}_{i}$:
\begin{linenomath}
\small
\begin{align}
p(R_{u i} | O_{u i}, \mathbf{x}_{u}, \mathbf{x}_{i})=p(R_{ui}|\mathbf{x}_u,\mathbf{x}_i).
\end{align}
\end{linenomath}
We observe that this assumption is actually referred to as  unconfoundedness assumption~\cite{rubin1974estimating} in causal inference: assuming that there are no other latent variables except the features that affect both the outcome and the treatment assignment. With this assumption, we  only have the distribution shift with respect to the exposure (propensity) probability (see Eq.~\eqref{Eq;shift}) and the conditional distribution shift  between $p(R_{ui}|\mathbf{x}_u,\mathbf{x}_i)$ and $p(R_{u i} | O_{u i}, \mathbf{x}_{u}, \mathbf{x}_{i})$ vanishes (i.e., labeling function $g(\mathbf{x}_{u},\mathbf{x}_{i})=k(\mathbf{x}_{u},\mathbf{x}_{i})$). Thus, these methods~\cite{schnabel2016recommendations,saito2020unbiased,wang2019doubly,zhu2020unbiased} can conduct 
unbiased estimation by inversely re-weighting each sample:
\begin{linenomath}
\small
\begin{align}
\widehat{\mathcal{L}}_{w}(f)=\frac{1}{N}
\sum\nolimits_{(\mathbf{x}_{u}, \mathbf{x}_{i},Y_{ui}) \in \mathcal{D}_{p}}\frac{1}{p(O_{ui}|\mathbf{x}_{u},\mathbf{x}_{i})} \ell(f({\mathbf{x}_{u}, \mathbf{x}_{i}}), Y_{ui}). \label{Eq:re-weight}
\end{align}
\end{linenomath}

It is straightforward to verify that $\widehat{\mathcal{L}}_{w}$ is an unbiased estimation of ideal risk: $\mathbb{E}_{P}[\widehat{\mathcal{L}}_{w}(f)]=\mathcal{L}_{Q}(f)$. Clearly, this
re-weighting objective can theoretically correct for the distribution shift caused by the exposure if $p(O_{ui}=1|\mathbf{x}_{u},\mathbf{x}_{i})$ is known in advance. Note that DR ~\cite{wang2019doubly} is also built on this re-weighting objective although it has an additional imputation model. We just focus on the re-weighting part of it here. While this objective has theoretical guarantee~\cite{wang2019doubly,saito2020unbiased,schnabel2016recommendations}, there are three crucial directions for improvement:

(1) The unconfoundedness assumption may not be true and cannot be examined in real recommendation scenarios~\cite{xu2020adversarial,ding2022addressing}, unless we can include every single factor that may affect users' decision-making process as a feature. However, there are other unobserved confounders, such as user social influence, item popularity effect, and public opinions, that cannot be captured through features. For example, as demonstrated in~\cite{krishnan2014methodology}, user ratings exhibit different distributions when users rate items before or after reading public opinions. Additionally, due to privacy restrictions, recommender systems inevitably face unobserved confounders. For instance, user financial status directly affects feedback but is not measurable in many recommender systems. Ignoring such confounders leads to an over-recommendation of inexpensive items. Nevertheless, current methods~\cite{wang2019doubly,saito2020unbiased,schnabel2016recommendations} do not consider these unobserved confounders.

(2) The theoretical analysis of this re-weighting objective~\cite{schnabel2016recommendations,wang2019doubly,saito2020unbiased} cannot explain and generalize well to many unbiased algorithms, especially those~\cite{bonner2018causal,DBLP:conf/sigir/ChenDQ0XCLY21,wang2021combating,liu2020general} that utilize unbiased uniform data.

(3) This objective also requires accurate estimation of the exposure probability, which is usually challenging~\cite{saito2020asymmetric,xu2020adversarial} and suffers from significant variance. Consequently, it performs poorly in empirical comparison to recent algorithms~\cite{DBLP:conf/sigir/ChenDQ0XCLY21,wang2020information,liu2020general}.

\section{Theoretical  Analysis}
\label{sec:theory}
In this section,  we first present our framework on unbiased recommendation from the distribution shift perspective with feature adaptation and derive two finite-sample generalization bounds. We provide a key insight that our theoretical framework is able to unify a series of recent unbiased learning objectives~\cite{bonner2018causal,liu2020general,wang2020information,DBLP:conf/sigir/ChenDQ0XCLY21,wang2021combating,liu2021mitigating}.

\subsection{Unbiased Learning via Feature Adaptation}
In this subsection, we show how feature adaptation is related to unbiased recommendation. Recall that we have logged feedback $\mathcal{D}_{P}$ from distribution $ P\left(\mathbf{x}_{u},\mathbf{x}_{i},O_{ui}\right)p(R_{ui}| O_{ui},\mathbf{x}_{u},\mathbf{x}_{i})$, where $P \triangleq P(\mathbf{x}_{u}, \mathbf{x}_{i},O_{ui})$ is the training marginal distribution over features. Similarly, we have the testing marginal distribution $Q\triangleq p(\mathbf{x}_{u})p(\mathbf{x}_{i})p(O_{ui})={1}/{|\mathcal{U}||\mathcal{I}|}$, meaning $(\mathbf{x}_{u},\mathbf{x}_{i})$ is sampled i.i.d. from uniform exposure distribution. Our goal is to learn a  function $f(\mathbf{x}_{u},\mathbf{x}_{i})$ which can approximate the optimal function $g(\mathbf{x}_{u}, \mathbf{x}_{i})$ which depends on preference distribution $p(R_{ui}|\mathbf{x}_{u},\mathbf{x}_{i})$.

To show how recent unbiased algorithms~\cite{liu2020general,DBLP:conf/sigir/ChenDQ0XCLY21,wang2020information,bonner2018causal} are related to feature adaptation, without loss of generality, we further consider the hypothesis $f(\mathbf{x}_{u},\mathbf{x}_{i})$ , which is composed of a two parts: $f=h \circ \phi$ where  $\phi \in$ $\Phi \subset\{\phi: \mathcal{X}_{u} \times \mathcal{X}_{i} \rightarrow \mathcal{Z}\}$ is the feature mapping function and  $h \in \mathcal{H} \subset\{h: \mathcal{Z} \rightarrow \mathcal{Y}\}$ is the hypothesis of the classification head. In general, $h$ is a linear or feed-forward network predictor. Given this, we notice that \citeauthor{ben2007analysis}~\cite{ben2007analysis} and \citeauthor{blitzer2007learning}~\cite{blitzer2007learning} proved the following bound on the unbiased risk $\mathcal{L}_{Q}(h \circ \phi)$ in terms of the empirical biased risk $\widehat{\mathcal{L}}_{P}(h \circ \phi)$ and the discrepancy between the training and testing distributions:

\begin{theorem}
\label{theorem:ben}
\cite{ben2007analysis,blitzer2007learning} Let $\mathcal{H}$ be a hypothesis space with VC-dimension $d$. ${{P}}(\mathbf{z}_{ui})$ (resp. ${{Q}}(\mathbf{z}_{ui})$) is the distribution over $\mathcal{Z}$ induced by marginal distribution $P(\mathbf{x}_{u}, \mathbf{x}_{i},O_{ui})$ (resp. $Q(\mathbf{x_u}, \mathbf{x}_{i},O_{ui})$) and $\phi$. Then, with probability (w.p)  at least $1-\delta$ over the natural exponential $e$, $\forall h \in \mathcal{H}$:
\begin{linenomath}
\small
\begin{align}
\mathcal{L}_{Q}(h \circ \phi) &\leq  \mathcal{\widehat{L}}_{P}(h \circ \phi)+\frac{1}{2} d_{\mathcal{H} \Delta \mathcal{H}}({{P}}(\mathbf{z}_{ui}) {{Q}}(\mathbf{z}_{ui}))\nonumber \\
&+\lambda(\phi) +\sqrt{\frac{4}{N} (d \log \frac{2 e N}{d}+\log \frac{4}{\delta})}, \text{where}
\end{align}
\end{linenomath}
$d_{\mathcal{H} \Delta \mathcal{H}}({{P}}(\mathbf{z}_{ui}), {{Q}}(\mathbf{z}_{ui}))=2  \sup _{h, h^{\prime} \in \mathcal{H}} \big|\mathbb{E}_{P(\mathbf{z}_{ui})}[\ell(h(\mathbf{z}_{ui}), h'(\mathbf{z}_{ui}))]-\mathbb{E}_{Q(\mathbf{z}_{ui})}[\ell(h(\mathbf{z}_{ui}),h'(\mathbf{z}_{ui}))]\big|$
is the $\mathcal{H} \Delta \mathcal{H}$-divergence~\cite{blitzer2007learning} which measures the discrepancy between two distributions on symmetric difference hypothesis space and $\lambda(\phi)=\inf _{h \in \mathcal{H}}(\mathcal{L}_{P}(h \circ \phi)+\mathcal{L}_{Q}(h \circ \phi))$ is the combined risk of the ideal hypothesis.
\end{theorem}
Theorem~\ref{theorem:ben} shows that the ideal risk $\mathcal{L}_{Q}(h \circ \phi)$ depends on three terms, which include the empirical risk $\mathcal{\widehat{L}}_{P}(h \circ \phi)$, the divergence between $P(\mathbf{z}_{ui})$ and $Q(\mathbf{z}_{ui})$, and the combined risk $\lambda(\phi)$. This bound serves as the theoretical foundation and has inspired the feature adaptation methods~\cite{ganin2015unsupervised,shrivastava2017learning}, which simultaneously minimizes the divergence between
$P(\mathbf{z}_{ui})$ and $Q(\mathbf{z}_{ui})$, and loss $\mathcal{\widehat{L}}_{P}(h \circ \phi)$.



This bound has made influential impacts in domain adaptation and we find there are two crucial directions to improve it for unbiased recommendation: (1) This bound considers aligning marginal distribution between $P(\mathbf{x}_{u},\mathbf{x}_{i},O_{ui})$ and $Q(\mathbf{x}_{u}, \mathbf{x}_{i},O_{ui})$ by using latent feature adaptation, however it does not theoretically reflect the unjustifiable unconfoundedness assumption as shown in \S~\ref{sec:assump}, namely the conditional distribution shift. This will make the upper bound loose when the unconfoundedness assumption is violated in the real-world. (2) This bound still can not give the guidance and explanation for unbiased learning objectives~\cite{bonner2018causal,liu2020general,DBLP:conf/sigir/ChenDQ0XCLY21,wang2021combating} that utilize unbiased uniform data. In what follows, we will introduce two  generalization bounds to provide interpretations for these learning objectives based on multi-task learning and meta-learning.

\subsection{Unbiased Multi-Task Learning Bound}
\label{subsec:concept}
\label{subsec:concept}
In this subsection, we give an unbiased multi-task  learning bound which measures the unconfoundedness assumption. We also demonstrate that a a series of existing unbiased recommendation algorithms~\cite{bonner2018causal,liu2020general,wang2020information,liu2021mitigating,saito2020unbiased,schnabel2016recommendations} including those using unbiased uniform data can be interpreted by this bound.

Specifically, some recent algorithms~\cite{bonner2018causal,liu2020general} conduct de-biasing learning via unbiased uniform data, which is collected by a random exposure probability $Q$ and
can reflect user preferences in an unbiased way. Thus, besides the biased data $\mathcal{D}_{P}$, we assume that we have some unbiased uniform data $\mathcal{D}_{Q}=\{\mathbf{x}_{u},\mathbf{x}_{i},Y_{ui}| O_{u, i}=1\}$ and the number of samples is $M$. Given the combined biased and unbiased data, these  algorithms~\cite{bonner2018causal,liu2020general,wang2020information,liu2021mitigating,saito2020unbiased,schnabel2016recommendations} generally have the following empirical multi-task learning objective:
\begin{linenomath}
\small
\begin{align}
\rho \mathcal{\widehat{L}}_{P}(h \circ \phi )+(1-\rho)\mathcal{\widehat{L}}_{Q}(h \circ \phi)+ \alpha R(\widehat{P}(\mathbf{z}_{ui}),\widehat{Q}(\mathbf{z}_{ui})), \label{Eq:multi-task}
\end{align}
\end{linenomath}
where $\rho\in[0,1]$. $\rho=1$ means that we do not have unbiased uniform data $\mathcal{D}_{Q}$. Thus, this formulation can  unify those algorithms~\cite{wang2020information,liu2021mitigating,schnabel2016recommendations} without using unbiased data. $R$ is the regularization function, and $\widehat{P}(\mathbf{z}_{ui})$, and $\widehat{Q}(\mathbf{z}_{ui})$ are empirical distributions of latent features over $P$ and $Q$, respectively. $\alpha$ is the hyper-parameters. $\mathcal{\widehat{L}}_{Q}(h \circ \phi )=\frac{1}{M} \sum_{(\mathbf{x}_{u}, \mathbf{x}_{i}, Y_{u i}) \in \mathcal{D}_{Q}} \ell(h \circ \phi(\mathbf{x}_{u}, \mathbf{z}_{i}), Y_{u i})$ is the empirical objective under unbiased uniform data $\mathcal{D}_{Q}$.  Based on this, we provide the following generalization bound:
\begin{theorem}
\label{theorem:multi-task}
Let $\mathcal{H}$ be a hypothesis space with VC-dimension d, and ${{P}}(\mathbf{z}_{ui})$ (resp. ${{Q}}(\mathbf{z}_{ui})$ is the probability density functions over $\mathcal{Z}$ induced by $P(\mathbf{x}_u, \mathbf{x}_{i},O_{ui})$ (resp. $Q(\mathbf{x}_{u},\mathbf{x}_{i},O_{ui})$) and $\phi$. $\tilde{g}$ (resp. $\tilde{k}$) is the labeling function over $\mathcal{Z}$ induced by $g$ (resp. $k$) and $\phi$. Then,  w.p.  at least  $1-\delta$ over the exponential $e$, $\forall h \in \mathcal{H}$:
\begin{linenomath} \small
\begin{align}
&\mathcal{{L}}_{Q}(h  \circ \phi) \leq  \rho \mathcal{\widehat{L}}_{P}(h \circ \phi )+(1-\rho)\mathcal{\widehat{L}}_{Q}(h \circ \phi) +\frac{\rho}{2}  d_{\mathcal{H} \Delta \mathcal{H}}({{P}}(\mathbf{z}_{ui}), {{Q}}(\mathbf{z}_{ui}))\nonumber \\
&+\rho\min \{\mathbb{E}_{P(\mathbf{z}_{ui})}[|\tilde{g}(\mathbf{z}_{ui})-\tilde{k}(\mathbf{z}_{ui})|], \mathbb{E}_{Q(\mathbf{z}_{ui})}[|\tilde{g}(\mathbf{z}_{ui})-\tilde{k}(\mathbf{z}_{ui})|]\}+ \nonumber \\
&(1-\rho) \sqrt{\frac{4}{M}(d \log \frac{2 e M}{d}+\log \frac{4}{\delta})}+\rho \sqrt{\frac{4}{N}(d \log \frac{2 e N}{d}+\log \frac{4}{\delta})}.
\end{align} 
\end{linenomath}
\end{theorem}

\noindent \textbf{Remark}. The proof is provided in Appendix~\ref{Appendix:multi-task}. This bound suggests that the ideal risk depends on the empirical multi-task learning error, the divergence of  feature distributions, and the distance $\min \{\mathbb{E}_{P(\mathbf{z}_{ui})}[|\tilde{g}(\mathbf{z}_{ui})-\tilde{k}(\mathbf{z}_{ui})|], \mathbb{E}_{Q(\mathbf{z}_{ui})}[|\tilde{g}(\mathbf{z}_{ui})-\tilde{k}(\mathbf{z}_{ui})|]\}$ of
labeling functions, which is essentially the divergence between conditional distributions $p(R_{u i} | O_{u i}, \mathbf{x}_{u}, \mathbf{x}_{i})$ and $p(R_{u i} | \mathbf{x}_{u}, \mathbf{x}_{i})$~\cite{zhao2019learning}.

Compared with the bound in Theorem~\ref{theorem:ben} and other  bounds in unbiased recommendation~\cite{schnabel2016recommendations,wang2019doubly}, the bound in Theorem~\ref{theorem:multi-task} has two key differences: (1) it involves a empirical multi-task learning objective. When $\rho=1$, the unbiased empirical error is not considered. With $\rho \in [0,1)$, we  introduce
both biased and unbiased data for de-biasing, and the generalizability of model could be improved. This appeals to us since we can theoretically justify algorithms, which employs unbiased data to conduct debiasing. (2) The term $\min \{\mathbb{E}_{P(\mathbf{z}_{ui})}[|\tilde{g}(\mathbf{z}_{ui})-\tilde{k}(\mathbf{z}_{ui})|], \mathbb{E}_{Q(\mathbf{z}_{ui})}[|\tilde{g}(\mathbf{z}_{ui})-\tilde{k}(\mathbf{z}_{ui})|]\}$   reflects the  unconfoundedness assumption. This bound explicitly considers this assumption and suggests that if it is violated, i.e., $p(R_{ui}|O_{ui},\mathbf{x}_{u},\mathbf{x}_{i})\neq p(R_{ui}|\mathbf{x}_{u},\mathbf{x}_{i})$, the bound will be loose. Thus, we should guarantee that the conditional distributions are not too far away from each other for successful unbiased recommendation.


The  bound in Theorem~\ref{theorem:multi-task}  enables us to interpret many  learning objectives ~\cite{schnabel2016recommendations,wang2020information,liu2020general,bonner2018causal,saito2020unbiased,wang2019doubly} in a unified perspective. Specifically, we show that they all fall into the multi-task objective in Eq.~\eqref{Eq:multi-task} and approximately minimize different terms in the bound. 

\vskip0.2em
\noindent \textbf{Re-weighting Objectives}~\cite{saito2020unbiased,schnabel2016recommendations}. These methods fall into the multi-task learning objective in Eq.~\eqref{Eq:multi-task} with $\rho=1$ and $\alpha=0$ since they do not utilize unbiased data and regularization. They re-weight the distribution $P$ via  $w(\mathbf{x}_{u},\mathbf{x}_{i})={1}/{p(O_{ui}|\mathbf{x}_{u},\mathbf{x}_{i})}$. By setting $\phi$ as the identity function, it is easy to verify that the first term in the bound becomes the re-weighting objective in Eq.~\eqref{Eq:re-weight}  and the third becomes $d_{\mathcal{H} \Delta \mathcal{H}}\big(w(\mathbf{x}_{u},\mathbf{x}_{i}){{P}}(\mathbf{x}_{u},\mathbf{x}_{i},O_{ui}), {{Q}}(\mathbf{x}_{u},\mathbf{x}_{i},O_{ui})\big)$ which equals to zero. Thus, they essentially minimize the first and third term in this bound with $\rho=1$ and $\phi$ being identity function. 
\vskip0.2em
\noindent \textbf{Information Bottleneck Objectives}~\cite{wang2020information,liu2021mitigating}. These algorithms also fall into the multi-learning objective with $\rho=1$ and $\alpha \neq 0$. The regularization term in Eq.~\eqref{Eq:multi-task} is instantiated as the information bottleneck to regularize the model to learn a invariant representation across training and testing distributions~\cite{wang2020information,liu2021mitigating}, which makes the $d_{\mathcal{H} \Delta \mathcal{H}}(P(\mathbf{z}_{ui}), {{Q}}(\mathbf{z}_{ui}))$ smaller. Thus, similar to  re-weighting objectives, essentially, information bottleneck objectives also  minimize the first and third term in our bound.

\vskip0.2em
\noindent \textbf{Multi-task Objectives}~\cite{liu2020general,bonner2018causal}. These algorithms utilize unbiased uniform data and have the regularization term for approximately reducing the divergence between $P(\mathbf{z}_{ui})$ and $Q(\mathbf{z}_{ui})$, thus $\rho \neq 1$ and $\alpha \neq 0$ in Eq.~\eqref{Eq:multi-task}. Specifically, $R$ is  $\|\mathbf{z}_{ui}-\mathbf{\hat{z}}_{ui}\|_{2}$ in \cite{bonner2018causal} where $\mathbf{z}_{ui}$ and $\mathbf{\hat{z}}_{ui}$ are sampled from $\widehat{P}(\mathbf{z}_{ui})$ and $\widehat{Q}(\mathbf{z}_{ui})$, respectively. \cite{liu2020general} designs other strategies for this regularization. Although the specific regularization  may be different, the high-level motivation of them can be theoretically understood as minimizing the first, second and approximately reducing the third divergence terms in this bound.

\subsection{Unbiased Meta-Learning Bound}
As an alternative, there are some algorithms such as Learning to Debias~\cite{wang2021combating} and AutoDebias~\cite{DBLP:conf/sigir/ChenDQ0XCLY21} utilizing the unbiased uniform data  via a meta-learning process. Typically, their objectives are still re-weighting objectives. However, unlike vanilla IPS or DR, they utilize the unbiased uniform data to train a weight function $w \in \mathcal{H}^{\prime} \subset\{w: \mathcal{X}_{u}\times \mathcal{X}_{i} \rightarrow \mathcal{W} \}$ such that the hypothesis $h$ trained on the biased data performs well on the unbiased data. Specifically, the meta-learning can be formulated as a bi-level optimization:
\begin{linenomath}
\small
\begin{align}
\min _{w} \widehat{\mathcal{L}}_{Q}(h(w)) ~\text {s.t.}~h(w)=\arg \min _{h} \widehat{\mathcal{L}}_{P_{w}}(h), \label{Eq:meta}
\end{align}
\end{linenomath}
where $P_{w}=w(\mathbf{x}_{u}, \mathbf{x}_{i})P(\mathbf{x}_{u},\mathbf{x}_{i},Y_{ui})$ stands for a new re-weighted distribution. $\widehat{\mathcal{L}}_{Q}(h^{*}({w})\circ \phi)$  is the upper-level objective under unbiased uniform data. $w$ and $h$ are optimized alternately until convergence. Note that, in these methods,  $h(w)$ is the function of re-weighting  and its hypothesis space $\mathcal{H}^{\prime}$ depends on biased training data due to the bi-level optimization~\cite{shu2019meta}. Empirically, this meta-learning objective perform well~\cite{DBLP:conf/sigir/ChenDQ0XCLY21,wang2021combating}. To theoretically understand them, we provide the following bound:
\begin{theorem}
\label{theorem:meta}
Let $\mathcal{H}$ and $\mathcal{H}^{\prime}$  be hypothesis spaces with VC-dimension $d$ and $d^{\prime}$, respectively. ${{P}}(\mathbf{z}_{ui})$ (resp. ${{Q}}(\mathbf{z}_{ui}))$ is the density functions over $\mathcal{Z}$ induced by $P(\mathbf{x}_u, \mathbf{x}_i,O_{ui})$ (resp. $Q(\mathbf{x}_u, \mathbf{x}_i,O_{ui})$) and $\phi$. $\tilde{g}$ (resp. $\tilde{k}$) is the latent labeling function induced by $g$ (resp. $k$) and $\phi$. Then w.p. at least $1-\sigma$ and natural exponential $e$, $\forall h \in \mathcal{H}$, we have:
\begin{linenomath}
\small
\begin{align}
&\mathcal{{L}}_{Q}(h\circ \phi) \leq  \rho \mathcal{\widehat{L}}_{P_{w}}(h \circ \phi )+(1-\rho)\mathcal{\widehat{L}}_{Q}(h(w)\circ \phi)    \\
&+\frac{\rho}{2} d_{\mathcal{H} \Delta \mathcal{H}}(w(\mathbf{x}_{u},\mathbf{x}_{i}){{P}}(\mathbf{z}_{ui}), {{Q}}(\mathbf{z}_{ui}))+\rho \sqrt{\frac{4}{N}(d \log \frac{2 e N}{d}+\log \frac{4}{\delta})}+ \nonumber \\
&+\rho\min \{\mathbb{E}_{P(\mathbf{z}_{ui})}[|\tilde{g}(\mathbf{z}_{ui})-\tilde{k}(\mathbf{z}_{ui})|], \mathbb{E}_{Q(\mathbf{z}_{ui})}[|\tilde{g}(\mathbf{z}_{ui})-\tilde{k}(\mathbf{z}_{ui})|]\} \nonumber \\
& +(1-\rho)(\frac{d' \log M-\log \delta}{3 M}+\sqrt{\frac{2(d' \log M-\log \delta)}{M}} ). \nonumber 
\end{align} 
\end{linenomath}
\end{theorem}
\noindent \textbf{Remark}. We provide the proof in Appendix~\ref{Appendix:meta}. This bound shows that the ideal risk  depends on four non-constant terms: the empirical training errors on biased and unbiased data, the discrepancy between latent feature distributions, and the distance between the conditional distribution. Unlike Theorem~\ref{theorem:multi-task}, this empirical error on the unbiased data is obtained via a meta validation process.

\noindent\textbf{Meta-learning Objectives}~\cite{DBLP:conf/sigir/ChenDQ0XCLY21,wang2021combating}. With the bound in Theorem~\ref{theorem:meta}, we can understand why recent meta-learning approaches can achieve good performance and provide interpretations for them. It is worth noting that the bi-level meta-learning objective in Eq.~\eqref{Eq:meta} exactly minimizes the first and second terms when $\phi$ is the identity function. Unlike re-weighting objectives, the $w(\mathbf{x}_{u},\mathbf{x}_{i})$  may not be the optimal sample weight, i.e., ${1}/{p(O_{ui}|\mathbf{x}_{u},\mathbf{x}_{i})}$. Thus, we can not theoretically guarantee that the third divergence term is small.  Meta-learning objectives also make the unconfoundedness assumptions and neglect the fourth term in the bound.

\section{Adversarial Self-Training}
\label{sec:Algorithm}
We have shown how our framework  allows us to reinterpret many learning objectives in unbiased recommendation. With the above theoretical analysis and insights, we summarize the limitations of current learning objectives as follows: (1) They all make the unconfoundedness assumption, namely they do not account for  the term about the conditional shifts in Theorems~\ref{theorem:multi-task} and \ref{theorem:meta}.  Nevertheless, the  unconfoundedness assumption is rarely true and can not be examined in the real-world~\cite{xu2020adversarial}.  (2) Some algorithms try to minimize the $\mathcal{H} \Delta \mathcal{H}$ divergence between marginal feature distributions via re-weighting~\cite{schnabel2016recommendations,saito2020unbiased} or different regularizers~\cite{wang2020information,liu2020general,bonner2018causal}. However re-weighting suffers from the variance issue~\cite{su2020doubly}. As for the regularizers~\cite{wang2020information,liu2020general,bonner2018causal}, they are only an approximation of the empirical $\mathcal{H} \Delta \mathcal{H}$-divergence which is hard to optimize.
(3) Meta-learning objectives need to compute the second-order gradient is expensive in both computational cost and memory~\cite{wang2021combating,DBLP:conf/sigir/ChenDQ0XCLY21,shu2019meta}.

To address these issues, we  exploit the theoretic analysis introduced in \S~\ref{sec:theory} to derive a practical algorithm, adversarial self-training, which can simultaneously alleviate the divergence of feature distributions and approximately account for unobserved confounders. We optimize a feature mapping such that the conditional distribution is invariant to the biased training and unbiased testing data.


\subsection{Adversarial Training for Adaptation}
Motivated by the discussion in \S~\ref{sec:theory}, we need to design a mechanism that enables feature adaptation for minimizing the $\mathcal{H} \Delta \mathcal{H}$-divergence. However it is   difficult to optimize it. Thus, we give a new generalization bound to guide the design of AST.
\begin{theorem}
\label{theorem:KL}
Let $\mathcal{H}$ be a hypothesis space with VC-dimension $d$. ${{P}}(\mathbf{z}_{ui})$ (resp. ${{Q}}(\mathbf{z}_{ui})$) is the distribution over $\mathcal{Z}$ induced by marginal distribution $P(\mathbf{x}_{u}, \mathbf{x}_{i},O_{ui})$ (resp. $Q(\mathbf{x_u}, \mathbf{x}_{i},O_{ui})$) and $\phi$. $\tilde{g}$ (resp. $\tilde{k}$) is the latent labeling function induced by $g$ (resp. $k$) and $\phi$ Then,  with probability at least $1-\delta$ over the natural exponential $e$, $\forall h \in \mathcal{H}$:
\begin{linenomath}
\small
\begin{align}
&\mathcal{{L}}_{Q}(h  \circ \phi) \leq  \rho \mathcal{\widehat{L}}_{P}(h \circ \phi )+(1-\rho)\mathcal{\widehat{L}}_{Q}(h \circ \phi) +\frac{\rho\sqrt{2 {\mathrm{KL}}({{P}}(\mathbf{z}_{ui}) \| {{Q}}(\mathbf{z}_{ui}))}}{2}   \nonumber \\
&+\rho\min \{\mathbb{E}_{P(\mathbf{z}_{ui})}[|\tilde{g}(\mathbf{z}_{ui})-\tilde{k}(\mathbf{z}_{ui})|], \mathbb{E}_{Q(\mathbf{z}_{ui})}[|\tilde{g}(\mathbf{z}_{ui})-\tilde{k}(\mathbf{z}_{ui})|]\}\nonumber\\
&+(1-\rho) \sqrt{\frac{4}{M}(d \log \frac{2 e M}{d}+\log \frac{4}{\delta})}+\rho \sqrt{\frac{4}{N}(d \log \frac{2 e N}{d}+\log \frac{4}{\delta})}. 
\end{align} 
\end{linenomath}
\end{theorem}
\noindent \textbf{Remark}. The proof is provided in Appendix~\ref{appendix:TheoremKL}. This bound provides theoretical justification for the use of KL (Kullback–Leibler)-divergence to conduct feature adaptation in unbiased recommendation. 
While  the explicit marginal densities of ${P}(\mathbf{z}_{ui})$ and ${Q}(\mathbf{z}_{ui})$ are intractable, we have data samples of them. This motivates us to leverage adversarial distribution matching strategies~\cite{nowozin2016f} to minimize  KL-divergence through a mini-max game with samples.  In particular, we  minimize ${\mathrm{KL}}({{P}(\mathbf{z}_{ui})\|{Q}(\mathbf{z}_{ui})})$ via the use of a critic function (the max-step), and then update the feature mapping $\phi$  accordingly to reduce the KL-divergence (the min-step). In this paper, we consider the
Fenchel-dual form of the KL-divergence~\cite{nowozin2016f}, i.e.,
\begin{linenomath}
\small
\begin{align}
{\mathrm{KL}}(P\| Q)=\mathbb{E}_{P}[\log P-\log Q]=\max _{\nu>0}\{\mathbb{E}_{P}[\log \nu]-\mathbb{E}_{Q}[\nu]+1\}.
\end{align}
\end{linenomath}
To optimize this Fenchel-dual form  in practice, we model $\log \nu$ using another function ${\theta}(\mathbf{z}_{ui})$ as our critic function. This results in the following adversarial neural estimator of ${\mathrm{KL}({{P}}(\mathbf{z}_{ui}) \| {{Q}}(\mathbf{z}_{ui}))}$:
\begin{linenomath}
\small
\begin{align}
&\widehat{\mathcal{L}}_{A}(\phi,\theta)=\min_{{\phi}}\max_{{\theta}}\mathbb{E}_{\mathbf{z}_{ui}={{\phi}}(\mathbf{x}_{u},\mathbf{x}_{i}),(\mathbf{x}_{u},\mathbf{x}_{i})\sim P(\mathbf{x}_{u}, \mathbf{x}_{i},O_{ui})}[{{\theta}}(\mathbf{z}_{ui})]\nonumber \\
&-\mathbb{E}_{\mathbf{z}_{ui}={{\phi}}(\mathbf{x}_{u},\mathbf{x}_{i}),(\mathbf{x}_{u},\mathbf{x}_{i})\sim Q(\mathbf{x}_{u}, \mathbf{x}_{i},O_{ui})}[\exp({{\theta}}(\mathbf{z}_{ui}))]\label{Eq:Adv}.
\end{align}
\end{linenomath}
Compared to $\mathcal{H} \Delta \mathcal{H}$-divergence, this is a much easier implementation to bound the ideal risk in Theorem~\ref{theorem:KL}.

 \begin{figure}[t]
\centering
  \subfigure{
    \includegraphics[width=0.45\textwidth]{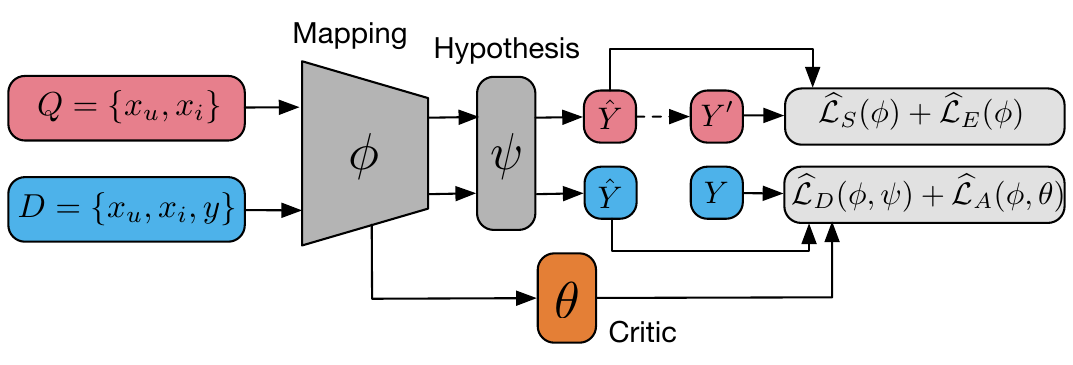}}
        \vskip -2em
     \caption{\small The architecture of AST. $\mathcal{D}$ is logged feedback, and $Q$ is randomly sampled features. Self-training and entropy minimization losses only update the feature mapping $\phi$ and does not update the prediction head $\psi$ since these two losses focus on searching for a feature mapping $\phi$ such that the conditional distribution is invariant to training and testing: $\mathbb{E}_{P}[Y_{ui}| \phi (\mathbf{x}_{u},\mathbf{x}_{i})]=\mathbb{E}_{Q}[Y_{ui}| \phi (\mathbf{x}_{u},\mathbf{x}_{i})]$.}\label{fig:model}
      \vskip -1.5em
\end{figure}
\subsection{Supervised Learning and Self-Training}
As suggested by the generalization bound in Theorem~\ref{theorem:KL}, we also need  to minimize the empirical learning error and the distance  between the optimal labeling functions. For the empirical multi-task learning error, we can directly minimize it by parameterizing
 hypothesis $h$ with function $\psi$:
\begin{linenomath}
\small
\begin{align}
&\mathcal{\widehat{L}}_{D}(\phi,\psi)=\mathbb{E}_{\mathbf{z}_{ui}={{\phi}}(\mathbf{x}_{u},\mathbf{x}_{i}),(\mathbf{x}_{u},\mathbf{x}_{i},y)\sim \mathcal{D}} [\ell ({\psi}(\mathbf{z}_{ui}), Y_{ui})], \label{Eq:sl}
\end{align}
\end{linenomath}
where $\mathcal{D}= \mathcal{D}_{P} \cup \mathcal{D}_{Q}$ is the whole set of data, including the biased data and the unbiased uniform data. Note that our algorithm can conduct de-biasing learning without unbiased uniform data when $\mathcal{D}= \mathcal{D}_{P}$. To further minimize the distance between conditional distributions (i.e., the regularizing term on conditional distributions), we need to search for a feature mapping $\phi$  such that the conditional distribution is invariant to training and testing: $\mathbb{E}_{P}[Y_{ui}| \phi (\mathbf{x}_{u},\mathbf{x}_{i})]=\mathbb{E}_{Q}[Y_{ui}| \phi (\mathbf{x}_{u},\mathbf{x}_{i})]$. If we have a  small amount of unbiased uniform data from $Q$, we can directly minimize this regularizing term on conditional distributions by jointly minimizing $\mathcal{\widehat{L}}_{D}(\phi,\psi)$ on both biased data and unbiased uniform data. 

However, in some scenarios, collecting unbiased uniform data is extraordinarily expensive~\cite{wang2020information,liu2020general}. Thus directly optimizing this term $\mathcal{\widehat{L}}_{D}(\phi,\psi)$ with unbiased uniform data  becomes inaccessible. To account for  this scenarios, in this paper, we propose to approximately evaluate and minimize this term by using self-training. 
Previous works~\cite{chen2020self,wei2020theoretical} have theoretically shown that self-training can learn the invariant predictive distribution, which can yield equally optimal performance across environments. This matches our goal of making conditional distribution invariant to the training and testing. Specifically, we adopt the principle of self-training, which has shown to be effective in semi-supervised learning~\cite{grandvalet2004semi,berthelot2019mixmatch,wei2020theoretical}.   Self-training first trains the feature mapping $\phi$ and prediction head $\psi$  via $\mathcal{\widehat{L}}_{D}(\phi,\psi)$ in Eq.~(\ref{Eq:sl}), and the trained model generates  pseudo-labels for the unlabeled data sampled from $Q(\mathbf{x}_{u},\mathbf{x}_{i}, O_{ui})$. Then self-training trains feature mapping with pseudo-labels as:
\begin{linenomath}
\small
\begin{align}
\mathcal{\widehat{L}}_{S}(\phi)= \mathbb{E}_{\mathbf{z}_{ui}={{\phi}}(\mathbf{x}_{u},\mathbf{x}_{i}),(\mathbf{x}_{u},\mathbf{x}_{i})\sim Q(\mathbf{x}_{u},\mathbf{x}_{i}, O_{ui})}[ \ell({\psi}(\mathbf{z}_{ui}), {Y'_{ui}})],
\end{align}
\end{linenomath}
where $Y'_{ui}={\bar{\psi}}({{\bar{\phi}}}(\mathbf{x}_{u}, \mathbf{x}_{i}))$ is the generated \textit{soft} pseudo-label (it can be the ground-true label if we have a small amount of unbiased uniform data). $\bar{\psi}$ and $\bar{\phi}$ indict that we do not propagate gradients through computing the pseudo labels. We empirically found that this self-training can effectively brings conditional distributions closer even we do not have any unbiased uniform data. In addition, inspired by the recent work~\cite{chen2020self} which proves that entropy minimization has a similar effect as
self-training algorithm, we also explicitly minimize the  entropy on unlabeled uniform data:
\begin{linenomath}
\small
\begin{align}
\mathcal{\widehat{L}}_{E}(\phi)=\mathbb{E}_{\mathbf{z}_{ui}={{\phi}}({x}_{u},{x}_{i}),({x}_{u},{x}_{i})\sim Q(\mathbf{x}_{u},\mathbf{x}_{i}, O_{ui})} [\mathrm{H}(\sigma({\psi}(\mathbf{z}_{ui}))],
\end{align}
\end{linenomath}
where $\mathrm{H}(X)=-\sum_{i=1}^{n} p(x_{i}) \log  p(x_{i})$ is the entropy of $X$. Intuitively, by minimizing this entropy, we can effectively  encourage the prediction to be  low-entropy (i.e., high-confidence) on unlabeled data and the classifier’s decision boundary should not pass through high-density regions of the data distribution~\cite{berthelot2019mixmatch}. In summary, the overall objective function  of AST could be formulated as follows:
\begin{linenomath}
\small
\begin{align}
\mathcal{L}=\min_{{\phi},{\psi}}\max_{{\theta}}\mathcal{\widehat{L}}_{D}(\phi,\psi)+\alpha\mathcal{\widehat{L}}_{A}(\phi,\theta)+\beta \mathcal{\widehat{L}}_{S}(\phi)+\gamma \mathcal{\widehat{L}}_{E}({\phi},{\psi}),
\end{align}
\end{linenomath}
where $\alpha$, $\beta$ and $\gamma$ are trade-off hyper-parameters  controlling the contributions of different losses. 

\noindent \textbf{Overall algorithm}.
Our full algorithm, Adversarial Self-Training (AST) is illustrated in Figure~\ref{fig:model} and is given in Algorithm~\ref{alg}.  At each iteration, we sample mini-batches from biased labeled and unlabeled unbiased data. We generate the pseudo-labels for the unlabeled unbiased data by the current model. Then the model is further trained on the labeled biased data and pseudo-labeled unbiased data. The critic $\theta$ is optimized adversarially for minimizing the shift between biased training and unbiased test distributions.

\noindent \textbf{Complexity}. As shown in Figure~\ref{fig:model}, compared with  other unbiased learning algorithms~\cite{saito2020asymmetric,wang2020information,liu2020general}, we  introduce only one linear  additional head for the critic which reuses embeddings obtained from the encoder. This suggests that our AST only introduces very few parameters and the model complexity is at the same level as other unbiased learning algorithms~\cite{saito2020asymmetric,wang2020information,liu2020general,bonner2018causal}.

\begin{algorithm}[t]
     \caption{Adversarial Self-Training (AST)} 
     \label{alg}
      \textbf{Input:} The collected biased data $\mathcal{D}_{P}$, unbiased data $D_{Q}$ and parameters $\alpha$, $\beta$, $\gamma$. Learning rate $\eta$. Maximum steps $T$. \\
   
       \textbf{if} $\mathcal{D}_{Q}\neq \emptyset$:
      $\mathcal{D}=\mathcal{D}_{P} \cup \mathcal{D}_{Q}$  \-\   \-\  
      \textbf{else}: $\mathcal{D}=\mathcal{D}_{P}$
       \\
       \textbf{For} {n= 1, $\cdots$, $T$} \textbf{do}\\{
      \-\ \-\ \-\ \-\ Sample batches of   $(\mathbf{x}_{u}, \mathbf{x}_{i}) \in Q(\mathbf{x}_{u}, \mathbf{x}_{i},O_{ui})$ \\
      \-\ \-\ \-\ \-\ Generate pseudo-labels $Y_{ui}'$ for each sample:  $(\mathbf{x}_{u}, \mathbf{x}_{i}, Y_{ui}')$\\
      \-\ \-\ \-\ \-\  $(\phi_{n},\psi_{n}) \leftarrow (\phi_{n-1},\psi_{n-1})-\eta (\nabla_{\phi} \mathcal{L},\nabla_{\psi} \mathcal{L})$ ~
      \\
       \-\ \-\ \-\ \-\  $\theta_{n} \leftarrow \theta_{n-1}+\eta \nabla_{\theta} \mathcal{L}$ 
    	}\\
      \textbf{Return} $\theta, \psi$, $\phi$
    \end{algorithm}
    \vskip -2em
\section{EXPERIMENT}
\begin{table*}[t]
\center
\caption{Unbiased learning performance of different algorithms  with standard deviation over five runs. The best and second best performance  are marked with boldface and underline, respectively.}
\vskip -1.5em
\setlength{\tabcolsep}{1.9mm}{
\resizebox{1.0\textwidth}{!}{
\begin{tabular}{@{}c cccc c cccc c cccc c l@{}}
\toprule[1.0pt]
 \multirow{2.5}{*}{Algorithms}  & \multicolumn{4}{c}{ \textbf{{Yahoo}}}  & & \multicolumn{4}{c}{\textbf{{Coat}}} & &  \multicolumn{4}{c}{\textbf{{Goodreads}}} \\

  \cline{2-5}    \cline{7-10}  \cline{12-15} 
  
  \cline{2-5}    \cline{7-10}  \cline{12-15} 
  
    & \multicolumn{2}{c}{\textbf{MF}}  & \multicolumn{2}{c}{\textbf{NCF}} 
    
    & & \multicolumn{2}{c}{\textbf{MF}} &\multicolumn{2}{c}{\textbf{NCF}}   
    & & 
    \multicolumn{2}{c}{\textbf{MF}} & \multicolumn{2}{c}{\textbf{NCF}}  
    \\
    \cline{2-3}  \cline{4-5}     \cline{7-8}  \cline{9-10} \cline{12-13}  \cline{14-15} 
     &  HR@5  & NDCG@5  & HR@5  & NDCG@5   & &   HR@5 & NDCG@5 & HR@5    & NDCG@5 &  &
     HR@5 & NDCG@5 & HR@5   & NDCG@5  &\\
     \toprule[1.0pt]
 {Biased}   &  $\underset{\color{black} {(\pm 0.0035)}}{0.6471}$ & $\underset{\color{black} {(\pm 0.0037})}{0.6542}$   & $\underset{\color{black} {(\pm 0.0029)}}{0.6352}$    & $\underset{\color{black} {(\pm 0.0017)}}{0.6584}$   & & $\underset{\color{black} {(\pm 0.0051)}}{0.4338}$     & $\underset{\color{black} {(\pm 0.0072)}}{0.6457}$   & $\underset{\color{black} {(\pm 0.0045)}}{0.4281}$     & $\underset{\color{black} {(\pm 0.0048)}}{0.6257}$   && $\underset{\color{black} {(\pm 0.0024)}}{0.3071}$ & $\underset{\color{black} {(\pm 0.0011)}}{0.1057}$  &  $\underset{\color{black} {(\pm 0.0029)}}{0.3214}$  &  $\underset{\color{black} {(\pm 0.0008)}}{0.1089}$ \\
      \toprule[1.0pt]
  IPS   & $\underset{\color{black} {(\pm 0.0047)}}{0.6598}$     & $\underset{\color{black} {(\pm 0.0052)}}{0.6661}$   &  $\underset{\color{black} {(\pm 0.0038)}}{0.6415}$   & $\underset{\color{black} {(\pm 0.0029)}}{0.6663}$    &    & $\underset{\color{black} {(\pm 0.0064)}}{0.4131}$     & $\underset{\color{black} {(\pm 0.0079)}}{0.6361}$  & $\underset{\color{black} {(\pm 0.0056)}}{0.4255}$    & $\underset{\color{black} {(\pm 0.0050)}}{0.6219}$  & & $\underset{\color{black} {(\pm 0.0038)}}{0.3156}$  & $\underset{\color{black} {(\pm 0.0027)}}{0.1108}$    & $\underset{\color{black} {(\pm 0.0041)}}{0.3462}$   & $\underset{\color{black} {(\pm 0.0018)}}{0.1152}$   \\
    DRJL   & $\underset{\color{black} {(\pm 0.0038)}}{0.6632}$   & $\underset{\color{black} {(\pm 0.0042)}}{0.6732}$    & $\underset{\color{black} {(\pm 0.0033)}}{0.6581}$   & $\underset{\color{black} {(\pm 0.0025)}}{0.6716}$    & & $\underset{\color{black} {(\pm 0.0040)}}{0.4255}$    & $\underset{\color{black} {(\pm 0.0049)}}{0.6378}$   & $\underset{\color{black} {(\pm 0.0023)}}{0.4391}$   & $\underset{\color{black} {(\pm 0.0027)}}{0.6381}$   && $\underset{\color{black} {(\pm 0.0034)}}{0.3237}$  & $\underset{\color{black} {(\pm 0.0021)}}{0.1255}$  &  $\underset{\color{black} {(\pm 0.0034)}}{0.3531}$  & $\underset{\color{black} {(\pm 0.0012)}}{0.1265}$    \\
      \toprule[1.0pt]
  CVIB   & $\underset{\color{black} {(\pm 0.0042)}}{0.6756}$   & $\underset{\color{black} {(\pm 0.0047)}}{0.6834}$   & $\underset{\color{black} {(\pm 0.0036)}}{\underline{{0.6635}}}$   & $\underset{\color{black} {(\pm 0.0027)}}{{0.6873}}$   & &  $\underset{\color{black} {(\pm 0.0039)}}{{0.4531}}$   & $\underset{\color{black} {(\pm 0.0034)}}{{ 0.6680}}$   & $\underset{\color{black} {(\pm 0.0029)}}{{0.4487}}$    & $\underset{\color{black} {(\pm 0.0033)}}{{0.6498}}$   &  & $\underset{\color{black} {(\pm 0.0025)}}{{0.3467}}$  & $\underset{\color{black} {(\pm 0.0026)}}{{0.1397}}$  & $\underset{\color{black} {(\pm 0.0038)}}{{0.3687}}$   & $\underset{\color{black} {(\pm 0.0015)}}{{ 0.1469}}$  \\
  ATT    & $\underset{\color{black} {(\pm 0.0044)}}{{0.6635}}$   & $\underset{\color{black} {(\pm 0.0049)}}{{0.6784}}$   & $\underset{\color{black} {(\pm 0.0037)}}{{ 0.6497}}$   & $\underset{\color{black} {(\pm 0.0023)}}{{0.6829}}$   & & $\underset{\color{black} {(\pm 0.0040)}}{{0.4371}}$    & $\underset{\color{black} {(\pm 0.0037)}}{{0.6349}}$    & $\underset{\color{black} {(\pm 0.0022)}}{{0.4357}}$    & $\underset{\color{black} {(\pm 0.0024)}}{{0.6358}}$    & & $\underset{\color{black} {(\pm 0.0035)}}{{0.3307}}$  & $\underset{\color{black} {(\pm 0.0026)}}{{0.1209}}$  & $\underset{\color{black} {(\pm 0.0030)}}{{0.3562}}$  & $\underset{\color{black} {(\pm 0.0014)}}{{ 0.1343}}$ \\
   ACL    & $\underset{\color{black} {(\pm 0.0040)}}{{0.6801}}$   & $\underset{\color{black} {(\pm 0.0045)}}{{0.6839}}$   & $\underset{\color{black} {(\pm 0.0032)}}{{0.6522}}$    & $\underset{\color{black} {(\pm 0.0022)}}{{0.6857}}$    & &  $\underset{\color{black} {(\pm 0.0036)}}{{0.4529}}$    & $\underset{\color{black} {(\pm 0.0033)}}{{\underline{0.6721}}}$   & $\underset{\color{black} {(\pm 0.0021)}}{{\underline{0.4631}}}$   & $\underset{\color{black} {(\pm 0.0029})}{{{0.6536}}}$  &  & $\underset{\color{black} {(\pm 0.0030)}}{{{0.3587}}}$  & $\underset{\color{black} {(\pm 0.0025)}}{{\underline{0.1477}}}$   & $\underset{\color{black} {(\pm 0.0034)}}{{{0.3714}}}$   &  $\underset{\color{black} {(\pm 0.0015)}}{{{0.1498}}}$   \\
     \toprule[1.0pt]
   KD    & $\underset{\color{black} {(\pm 0.0043)}}{{{0.6779}}}$   & $\underset{\color{black} { (\pm 0.0044)}}{{{0.6781}}}$   & $\underset{\color{black} {(\pm 0.0031)}}{{{0.6571}}}$    & $\underset{\color{black} {(\pm 0.0024)}}{{{0.6814 }}}$   & & $\underset{\color{black} {(\pm 0.0038)}}{{{0.4561}}}$   & $\underset{\color{black} {(\pm 0.0036)}}{{{0.6584}}}$   & $\underset{\color{black} {(\pm 0.0024)}}{{{0.4451}}}$    & $\underset{\color{black} {(\pm 0.0021)}}{{{0.6471}}}$    & &  $\underset{\color{black} {(\pm 0.0029)}}{{{0.3533}}}$  & $\underset{\color{black} {(\pm 0.0023)}}{{{0.1368}}}$  &  $\underset{\color{black} {(\pm 0.0035)}}{{{0.3669}}}$  &  $\underset{\color{black} {(\pm 0.0012)}}{{{0.1405}}}$  \\
AutoDebias    &  $\underset{\color{black} {(\pm 0.0046)}}{{\underline{0.6835}}}$   & $\underset{\color{black} {(\pm 0.0051)}}{{\underline{0.6959}}}$   & $\underset{\color{black} {(\pm 0.0035)}}{{{0.6609}}}$  & $\underset{\color{black} {(\pm 0.0028)}}{{\underline{0.6925}}}$    & &  $\underset{\color{black} {(\pm 0.0042)}}{{\underline{0.4628}}}$   & $\underset{\color{black} {(\pm 0.0037)}}{{{0.6651}}}$    &  $\underset{\color{black} {(\pm 0.0028)}}{{{0.4568}}}$    & $\underset{\color{black} {(\pm 0.0035)}}{{\underline{0.6587}}}$   & & $\underset{\color{black} {(\pm 0.0041)}}{{\underline{0.3608}}}$   & $\underset{\color{black} {(\pm 0.0026)}}{{{0.1428}}}$   & $\underset{\color{black} {(\pm 0.0038)}}{{\underline{0.3751}}}$   &  $\underset{\color{black} {(\pm 0.0016)}}{{\underline{0.1518}}}$    \\
\toprule[1.0pt]
 AST    & $\underset{\color{black}{(\pm 0.0041)}}{{{\textbf{0.6985}}}}$     & $\underset{\color{black}{(\pm 0.0046)}}{{{\textbf{0.7147}}}}$  & $\underset{\color{black}{(\pm 0.0030)}}{{{\textbf{0.6813}}}}$     & $\underset{\color{black}{(\pm 0.0021)}}{{{\textbf{0.7094}}}}$    & & $\underset{\color{black}{(\pm 0.0037)}}{{{\textbf{0.4775}}}}$   & $\underset{\color{black}{(\pm 0.0035)}}{{{\textbf{0.6819}}}}$    & $\underset{\color{black}{(\pm 0.0023)}}{{{\textbf{0.4728}}}}$     & $\underset{\color{black}{(\pm 0.0031)}}{{{\textbf{0.6630}}}}$    & & $\underset{\color{black}{(\pm 0.0036)}}{{{\textbf{0.3712}}}}$ & $\underset{\color{black}{(\pm 0.0024)}}{{{\textbf{0.1678}}}}$  & $\underset{\color{black}{(\pm 0.0033)}}{{{\textbf{0.3834}}}}$ &  $\underset{\color{black}{(\pm 0.0012)}}{{{\textbf{0.1655}}}}$  \\
  \bottomrule[1.0pt]
\end{tabular}}}
\vskip -1em
\label{Table:overall}
\end{table*}

In this section, we empirical evaluate the performance of the proposed AST~\footnote{The code of AST is available at \url{https://github.com/UnbiasedRS/AST}.} on both real-world and semi-synthetic datasets. 

\begin{table}[!t]
\caption{Ablation study (NDCG@5) with MF backbone.}
\centering
\vskip -1.5em
\begin{tabular}{lcccc l}
   \toprule[1.0pt]
\textbf{Methods}  & \textbf{{Coat}}  & \textbf{{Yahoo}} & \textbf{{Goodreads}}  \\
   \toprule[1.0pt]
AST w/o A & $\underset{\color{black} {(\pm 0.0039)}}{{0.6628}}$   & $\underset{\color{black} {(\pm 0.0043)}}{{0.7025}}$  & $\underset{\color{black} {(\pm 0.0039)}}{{0.1498}}$  \\
AST w/o S & $\underset{\color{black} {(\pm 0.0029)}}{{ 0.6693}}$ & $\underset{\color{black} {(\pm 0.0038)}}{\underline{0.7114}}$  & $\underset{\color{black} {(\pm 0.0022)}}{\underline{0.1615}}$   \\
AST w/o E & $\underset{\color{black} {(\pm 0.0041)}}{\underline{0.6733}}$  &  $\underset{\color{black} {(\pm 0.0052)}}{{0.7104}}$  & $\underset{\color{black} {(\pm 0.0031)}}{{ 0.1545}}$  \\
AST w/o S \& E & $\underset{\color{black} {(\pm 0.0032)}}{{0.6587}}$  &  $\underset{\color{black} {(\pm 0.0039)}}{{0.6978}}$  & $\underset{\color{black} {(\pm 0.0025)}}{{ 0.1317}}$  \\
\toprule
Biased  & $\underset{\color{black} {(\pm 0.0072)}}{0.6457}$  & $\underset{\color{black}{(\pm 0.0037})}{0.6542}$ & $\underset{\color{black} {(\pm 0.0011)}}{0.1057}$ \\
AutoDebias  & $\underset{\color{black} {(\pm 0.0037)}}{{{0.6651}}}$ & $\underset{\color{black} {(\pm 0.0051)}}{{\underline{0.6959}}}$ & $\underset{\color{black} {(\pm 0.0026)}}{{{0.1428}}}$ \\
AST  & $\underset{\color{black}{(\pm 0.0035)}}{{{\textbf{0.6819}}}}$ & $ \underset{\color{black}{(\pm 0.0046)}}{{{\textbf{0.7147}}}}$ & $\underset{\color{black}{(\pm 0.0024)}}{{{\textbf{0.1678}}}}$  \\
\toprule[1.0pt]
\end{tabular}\label{Ablation}
\vskip -1.5em
\end{table}

\subsection{Experimental Settings}
\subsubsection{Datasets} Following previous works~\cite{schnabel2016recommendations,saito2020unbiased,wang2021combating,wang2020information,liu2020general}, we use two real-world datasets: \textbf{{Yahoo}}
~\cite{marlin2007collaborative} and \textbf{{Coat}}~\cite{schnabel2016recommendations}. These two datasets are suitable for verifying our theoretical analysis and evaluating our AST since they contain both biased and unbiased data, where unbiased data is formed by randomly assigning items to users for ratings. Thus they can be used to measure the unbiased generalization performance with selection bias. \textbf{{Yahoo}}\footnote{\url{https://webscope.sandbox.yahoo.com/}}: Its biased training set has approximately 300,000 five-star ratings of 1,000 songs from 15,400 users. It collects an unbiased test set by asking 5,400 users to rate 10 randomly displayed songs. \textbf{{Coat}}\footnote{\url{https://www.cs.cornell.edu/~schnabts/mnar/}}: It has 290 users and 300 items. Each user rates 24 items by themselves forming 6,500 biased five-star ratings, and is asked to rate 16 uniformly displayed items as the unbiased set. Since these two real-world datasets are relatively small, we also generate a relatively large semi-synthetic dataset based on ~\textbf{{Goodreads}}\footnote{\url{https://sites.google.com/eng.ucsd.edu/ucsdbookgraph}}. It is a book recommendation dataset~\cite{wan2018item} and we use the book review subset in history and biography, containing 238,450 users, 302,346  items, and 2,066,193 five-star ratings.  

\subsubsection{Prepossessing}Following~\cite{wang2020information,liu2020general,DBLP:conf/sigir/ChenDQ0XCLY21}. For all datasets, we treat rating which is 3 or higher as positive feedback and the others as negative. For \text{{GoodReads}}, we remove those items and users that have less than 20 interactions.

\subsubsection{Splitting and Simulation Settings} For \text{{Coat}} and \text{{Yahoo}},
following~\cite{liu2020general,DBLP:conf/sigir/ChenDQ0XCLY21}, we treat all biased $\mathcal{D}_P$ data as training set, and split the unbiased data into three parts: $5\%$ as additional training set $\mathcal{D}_{Q}$ to help training, $5 \%$ as validation set, and the remaining $90 \%$ as test set. Since  \text{{Goodreads}} does not contain an unbiased testing set, we simulate a semi-synthetic dataset to facilitate
ground-truth evaluation against a fully known relevance and exposure parameter. Strictly following previous works~\cite{xu2020adversarial,saito2020unbiased}, we first hold out the last feedback of all users in the last time slice  as the test data and the feedback before the last is treated as the validation set. The remaining feedback serves as training set. We train a MF model to approximate the rating matrix by minimizing the mean-squared loss based on the training set. Then ground-truth preference probability for unbiased testing is $p(Y_{u,i}=1| O_{u, i}):=\sigma(\hat{\mathbb{E}}[R_{u, i}| O_{u, i}]+\epsilon_R)$  where $\hat{\mathbb{E}}[R_{u, i}| O_{u, i}]$ is the model output and $\epsilon_{R}$ is Gaussian noisy controlling  randomness of preference caused by unobserved confounders. Then, similar to~\cite{saito2020unbiased,xu2020adversarial}, we utilize another logistic MF predicting if the 
rating is observed as the exposure $\hat{p}(O_{u, i})$ model. The final log-exposure probability  $\log p(O_{u, i})=\log \hat{p}(O_{u, i})+\epsilon_{O}$, where $\epsilon_{O}$ measures  the extra randomness of exposure by unobserved confounders.  In our experiments,
we set $\epsilon_{O}$ and $\epsilon_{R}$ as five~\cite{saito2020unbiased}. Following the generative model in \S~\ref{Pre}, we generate the biased training feedback as $p(Y_{u, i}=1)=p(Y_{u, i}=1 |O_{u, i}) p(O_{u, i})$. With this simulation process, we can obtain the true relevance, exposure parameters and unobserved confounders for the unbiased evaluation.

\subsubsection{Setup} 
We compare AST with the following  learning objectives: direct supervised training (Biased), IPS~\cite{schnabel2016recommendations,saito2020unbiased}, DRJL~\cite{wang2019doubly}, CVIB~\cite{wang2020information}, ACL~\cite{xu2020adversarial}, ATT~\cite{saito2020asymmetric}, KD~\cite{liu2020general}, and AutoDebias~\cite{DBLP:conf/sigir/ChenDQ0XCLY21}. Since our AST is high-level learning approach that is compatible with almost all existing recommendation models, we consider two representative recommendation models: matrix factorization (MF)~\cite{koren2009matrix} and neural collaborative filtering (NCF)~\cite{he2017neural}.  Following previous works~\cite{wang2019doubly,xu2020adversarial,saito2020asymmetric}, we utilize Hit Ratio (HR)$@$5  and NDCG$@$5  to evaluate the unbiased ranking performance. For all methods, the hyper-parameter search space is: dropout \{0.2, 0.4, 0.6\}, learning rate \{0.001, 0.005, 0.01\},  weight-decay \{1e-4, 1e-5, 1e-6\}, embedding dimension \{64, 128, 256\}. Specifically, for  AST, we further search $\alpha$, $\beta$ and $\gamma$ from space \{0.2, 0.4, 0.6, 0.8\}. For a rigorous and fair comparison, we use the grid search to find the best hyperparameters 
of the baselines for all methods based on the validation performance. The detailed implementations of  baselines are given in Appendix~\ref{appendix:exp}.


 
  



\subsection{Unbiased Learning Performance}
\label{sec:overall}
Table~\ref{Table:overall} gives the unbiased learning results of AST and the baselines with NCF and MF as backbones, respectively. From the table, we observe: (\textbf{1}) As suggested by our theoretical analysis, AST  significantly outperforms other algorithms, which demonstrates its good generalization ability. This is because our AST can effectively minimize the generalization bound of the ideal risk; (\textbf{2})
Overall,  AST consistently outperforms other baselines  on all datasets with both MF and NCF backbones, which implies the effectiveness of AST and also shows the flexibility and robustness of AST for various backbones; (\textbf{3}) Although IPS and DR come with strong theoretical insights, they empirically perform poorly. In contrast,  AST is empirically effective  while maintaining  rigorous theoretical justification; (\textbf{4}) AST  outperforms baselines on \textsc{Goodreads} which has both selection bias and  unobserved exposure factors, demonstrating that
AST can simultaneously account for both selection bias and the
unconfoundedness assumption, which makes the  generalization  bound tighter; (\textbf{5}) As suggested by our theoretical analysis,  KD and AutoDebias can improve performance by utilizing the unbiased uniform data. However, as shown in Table~\ref{Table:overall}, our AST still outperforms them by a large margin. 

\begin{figure}[t]
\centering
      \vskip -0.5em
      \subfigure{
    \includegraphics[width=0.28\textwidth]{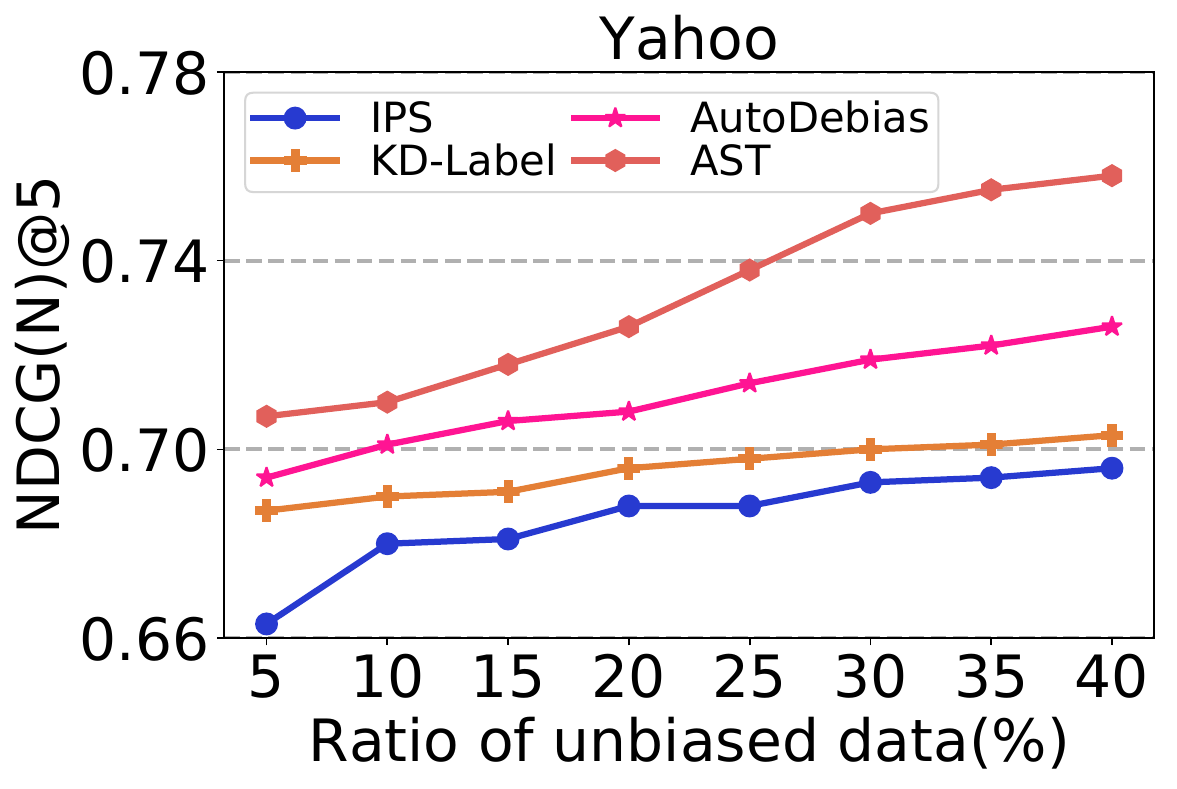}}
        \vskip -2em
     \caption{\small Comparison with different ratio of uniform data.}\label{fig:uniform-ratio}
    \vskip -1em
\end{figure}

We also evaluate the performance of AST and baselines under different percentage of unbiased uniform data, which is presented in Figure~\ref{fig:uniform-ratio}. We can find that AST significantly outperforms baselines deigned for using unbiased data under all the  data ration performed and the improvement increases when larger amount of uniform data is available. This is consistent with our theoretical results, since more unbiased data can make the generalization bound smaller. 

\subsection{Ablation Study and Parameter Sensitivity}
\textbf{Setup}. To take a deeper examination on how different components affect AST performance, we conduct ablation study and parameter sensitivity analysis. We follow the same setup in \S~\ref{sec:overall} and  build the following variants of AST: (i) Removing adversarial matching component (\text{AST w/o A}); (ii) Removing self-training  (\text{AST w/o S}); and (iii) Removing entropy minimization  (\text{AST w/o E}).\\
\textbf{Results}. Table~\ref{Ablation} shows the ablation study results. We can find that our all designed components can contribute to the performance gain, and their  contributions are complementary to each other.  
\begin{figure}[t]
\vskip -1em
\centering
  \subfigure{
    \includegraphics[width=0.22\textwidth]{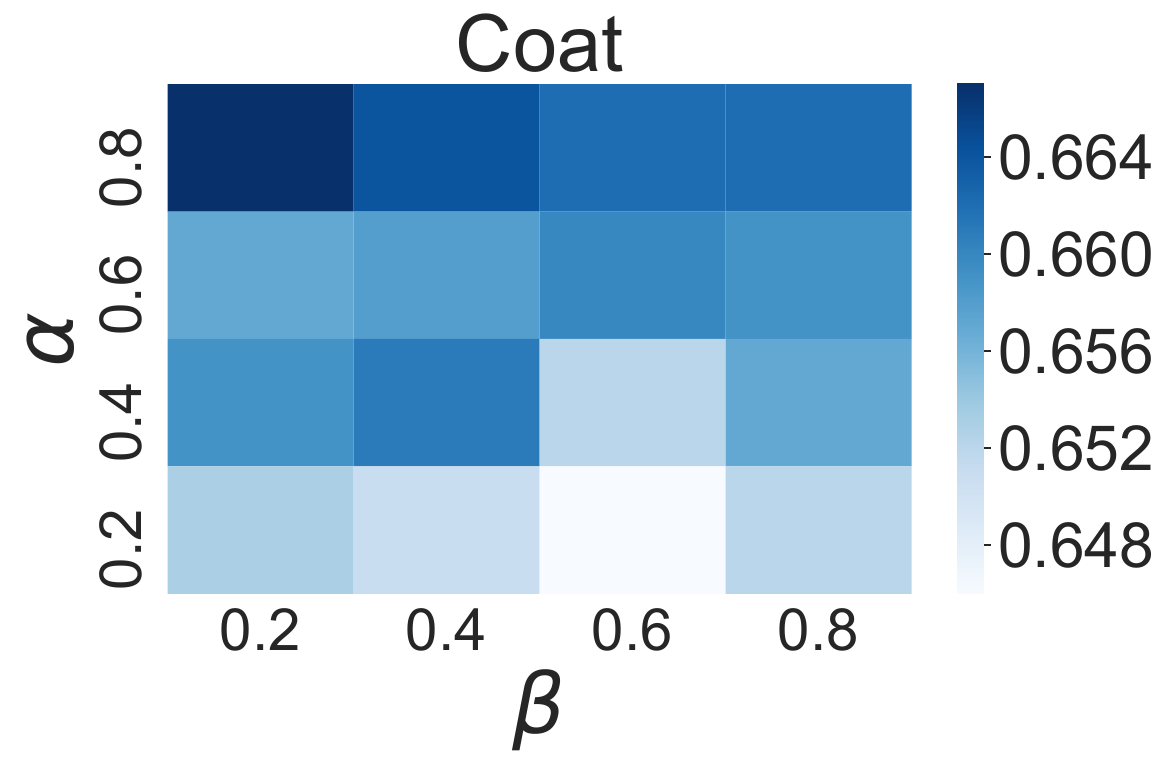}}
      ~~\subfigure{
    \includegraphics[width=0.22\textwidth]{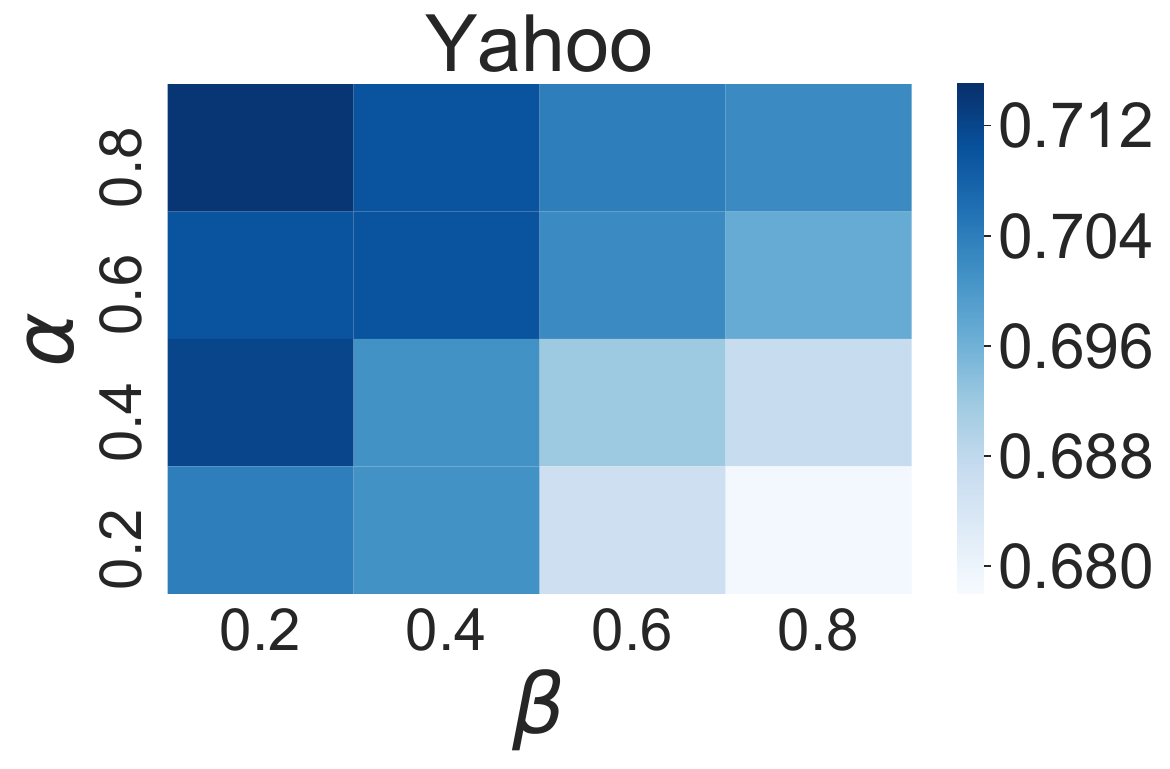}}
        \vskip -2.2em
     \caption{\small Sensitive analysis with NDCG$@$5 and MF backbone.}\label{fig:Sensitivity}
    \vskip -1.5em
\end{figure}
We also investigate the sensitivity of hyperparameters $\alpha$ and $\beta$, where  $\alpha$ and $\beta$ control the
contribution of adversarial matching and self-training, respectively. The $\gamma$ shares the same trend as $\beta$, thus we omit it to save space. We vary $\alpha$ and $\beta$ as $[0.2, 0.4, 0.6, 0.8]$ and report the results in Fig.~\ref{fig:Sensitivity}. We find that: (\textbf{i}) The performance of AST is generally better and stable 
when $\alpha \in \{0.6, 0.8\}$ and $\beta \in \{0.2, 0.6\}$, which eases the hyperparameter selection. (\textbf{ii}) We can balance the adversarial matching and self-training by varying $\alpha$ and $\beta$, leading  to better generalization  performance. This confirms the motivation of jointly alleviating selection bias and unobserved confounders.

\subsection{Performance on Challenging Scenarios}
In this subsection, we consider two more challenging scenarios. The first is de-biasing without any unbiased training data. This scenario is realistic since collecting unbiased  data is typically expensive. We compare AST with  baselines that do no need unbiased training data. Strictly following~\cite{liu2020general,DBLP:conf/sigir/ChenDQ0XCLY21},  we regard the biased data as training set and randomly sample $5\%$ ratings from the unbiased test data as the validation set. The results are reported in Table~\ref{Table:without}. We can find that AST can still achieve the best performance compared to baselines. In particular, our AST outperforms ATT and CVIB, which further shows the effectiveness of  adversarial feature adaptation via the KL-divergence (i.e, Theorem~\ref{theorem:KL}).  

The second  scenario is the implicit feedback. Compared to explicit feedback, the implicit feedback is much more challenging since we do not have any negative evidence in the learning~\cite{saito2020unbiased,xu2020adversarial}. Thus, we further evaluate AST on this scenario. To generate implicit feedback, we  use  Yahoo and Coat datasets but remove the negative feedback in the training data. Table~\ref{Table:implicit}  gives the results. We find that AST outperforms all baselines, which shows that AST can also effectively alleviate the selection bias issue for implicit feedback data. This matches our theoretical analysis since the ideal risk  still can be bounded under the setting of implicit feedback. 
\begin{figure}[t]
\centering
  \subfigure{
    \includegraphics[width=0.470\textwidth]{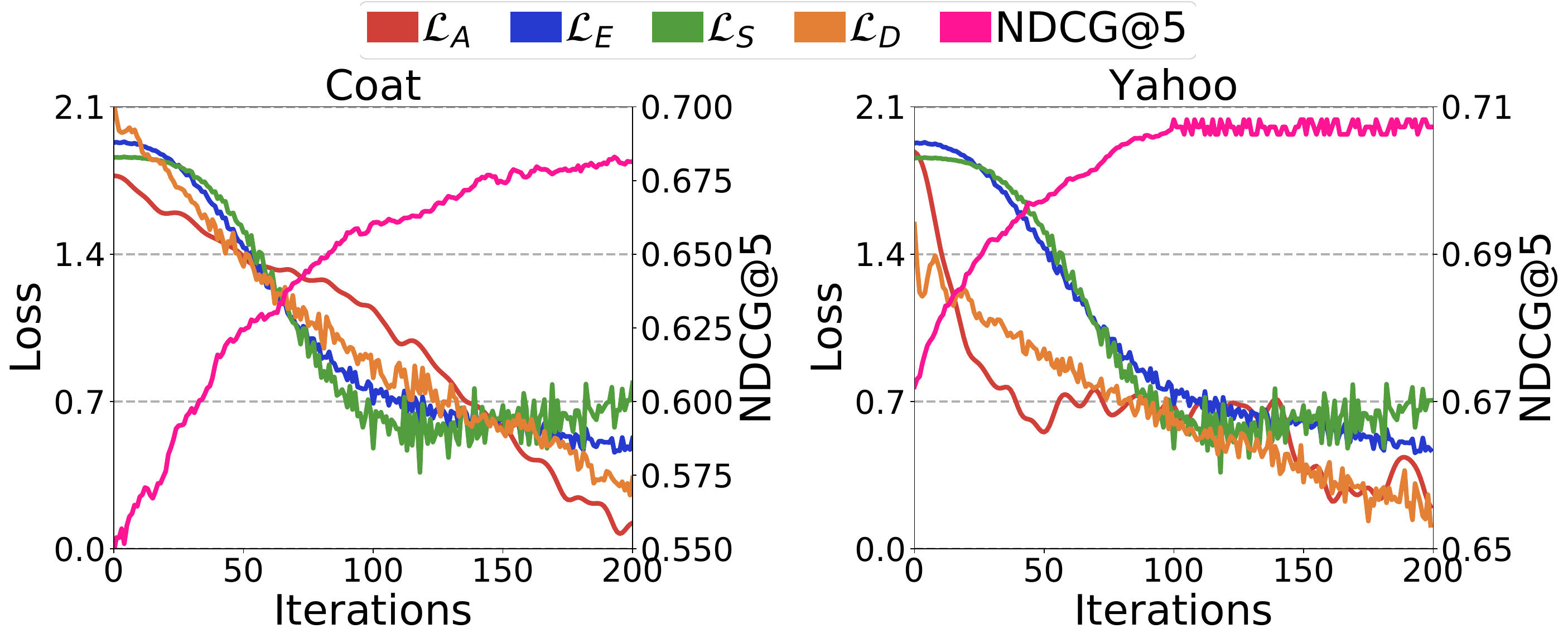}}
        \vskip -2em
     \caption{\small Generalization performance and training losses.}\label{fig:loss}
    \vskip -1.5em
\end{figure}

\begin{table}[t]
	\centering
	\caption{\small Performance (NDCG@5) without unbiased data.} 
	\vskip -1.5em
	\label{Table:without}
		\setlength{\tabcolsep}{3.5mm}{
		\resizebox{0.46\textwidth}{!}{
		\begin{tabular}{lcccccl}
			\toprule
			& \multicolumn{2}{c}{{ \textbf{Yahoo}}} & & \multicolumn{2}{c}{{\textbf{Coat} }} \\
			\Cline{2-3} \Cline{5-6}           & \multicolumn{1}{c}{\textbf{MF}} & \multicolumn{1}{c}{\textbf{NCF}} & & \multicolumn{1}{c}{\textbf{MF}} & \multicolumn{1}{c}{\textbf{NCF}} & \\
			   \toprule[1.0pt]
Biased & 0.6533 & 0.6714& & 0.6205 & 0.6330 & \\
    \toprule[1.0pt]
IPS &0.6661 & 0.6756 & & 0.6147 & 0.6440 &\\
DRJL & 0.6673 & 0.6789 && 0.6433 & 0.6376 &\\
    \toprule[1.0pt]
ATT & 0.6778  &  0.6788 && 0.6332 & 0.6472 &\\
CVIB & 0.6717 & 0.6906 && 0.6529 & 0.6519 &\\
 \toprule[1.0pt]
AST &   \textbf{0.6898} & \textbf{0.7004} && \textbf{0.6712} & \textbf{0.6589} &\\
			    \toprule[1.0pt]
		\end{tabular}}}
\end{table}%

\begin{table}[t]
	\centering
    \vskip -1em
	\caption{\small Performance (NDCG$@$5) on implicit feedback.} 
	\vskip -1.5em
	\label{Table:implicit}
		\setlength{\tabcolsep}{3.5mm}{
		\resizebox{0.48\textwidth}{!}{
		\begin{tabular}{lcccccl}
			\toprule
			& \multicolumn{2}{c}{{ \textbf{Yahoo}}} & & \multicolumn{2}{c}{{\textbf{Coat} }} \\
			\Cline{2-3} \Cline{5-6}           & \multicolumn{1}{c}{\textbf{MF}} & \multicolumn{1}{c}{\textbf{NCF}} & & \multicolumn{1}{c}{\textbf{MF}} & \multicolumn{1}{c}{\textbf{NCF}}& \\
			   \toprule[1.0pt]
Biased &0.6914 & 0.6233 & & 0.5514 & 0.6373& \\
    \toprule[1.0pt]
IPS &0.7011 & 0.6484 & & 0.5458 & 0.6144 &\\
DRJL & 0.7025 & 0.6517 && 0.5833 & 0.6181& \\
    \toprule[1.0pt]
ACL &0.7097  &  0.6885 && 0.5921 & 0.6348 &\\
KD &0.7152  &  0.6758 && 0.5692 & 0.6214 &\\
AutoDebias & 0.7195 & 0.6742 && 0.5873 & 0.6388& \\
 \toprule[1.0pt]
AST & \textbf{0.7248} & \textbf{0.7026} && \textbf{0.6037} & \textbf{0.6631}& \\
			    \toprule[1.0pt]
		\end{tabular}}}
  \vskip -1em
\end{table}%

\subsection{Deeper Understanding of AST}
\begin{figure}[t]
\centering
  \subfigure{
    \includegraphics[width=0.22\textwidth]{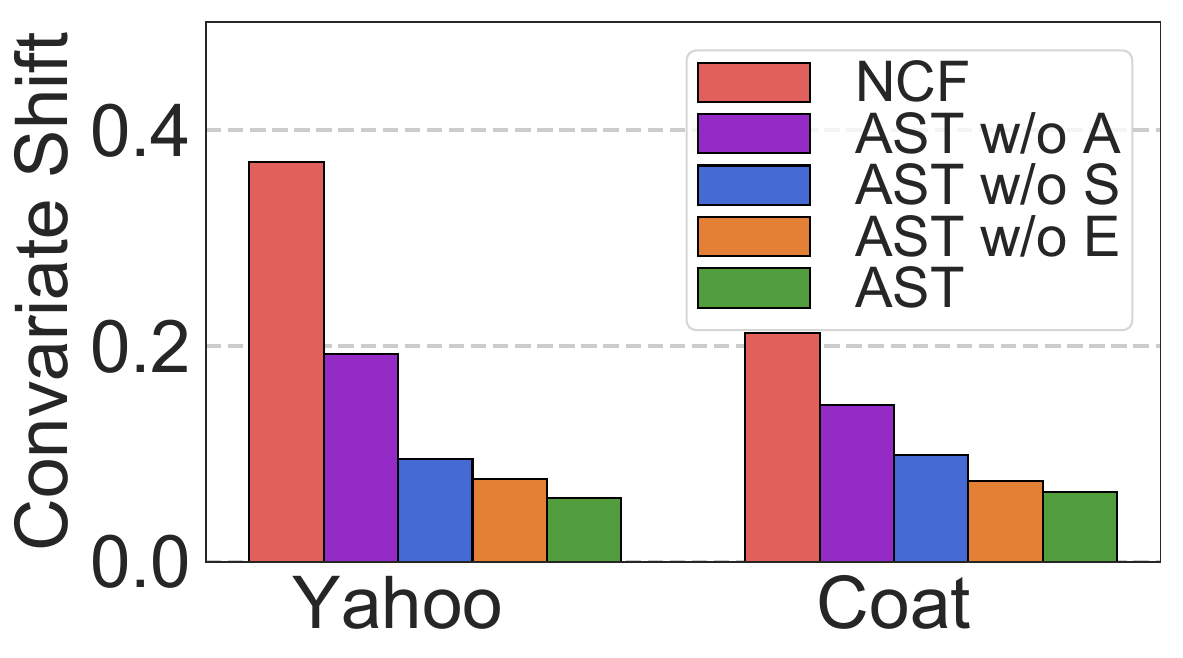}}
      \subfigure{
    \includegraphics[width=0.22\textwidth]{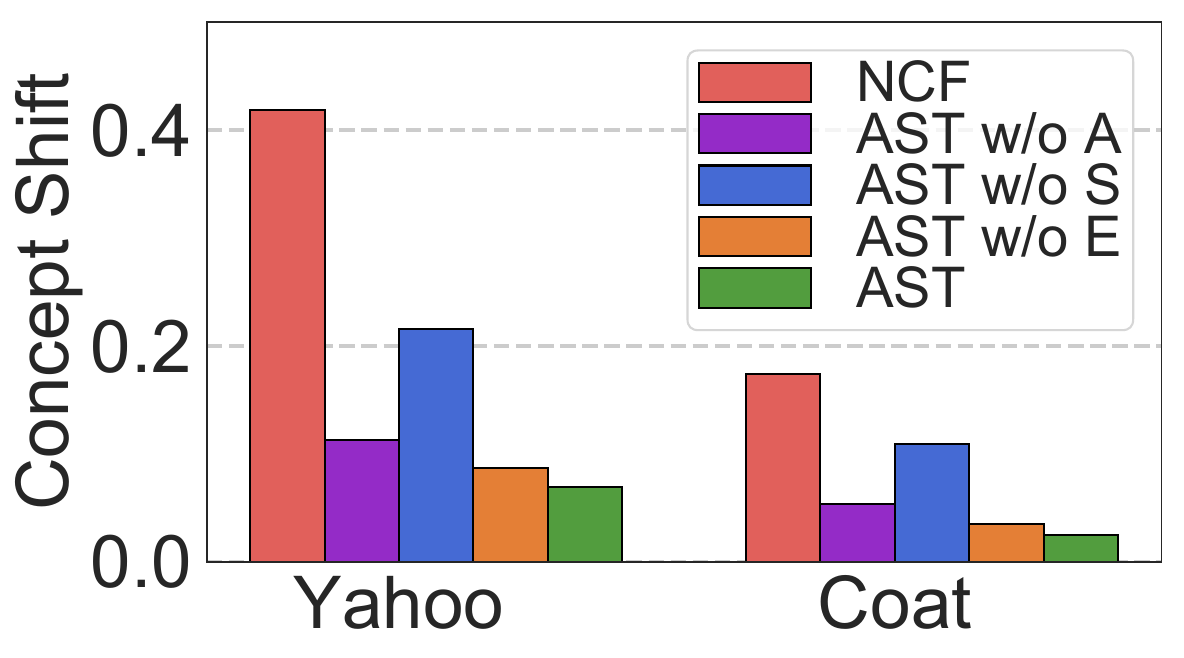}}
        \vskip -2em
     \caption{\small Marginal and conditional shifts on Yahoo and Coat.}\label{fig:distancNCF}
    \vskip -2em
\end{figure}

\begin{figure}[t]
\centering
  \subfigure{
    \includegraphics[width=0.38\textwidth]{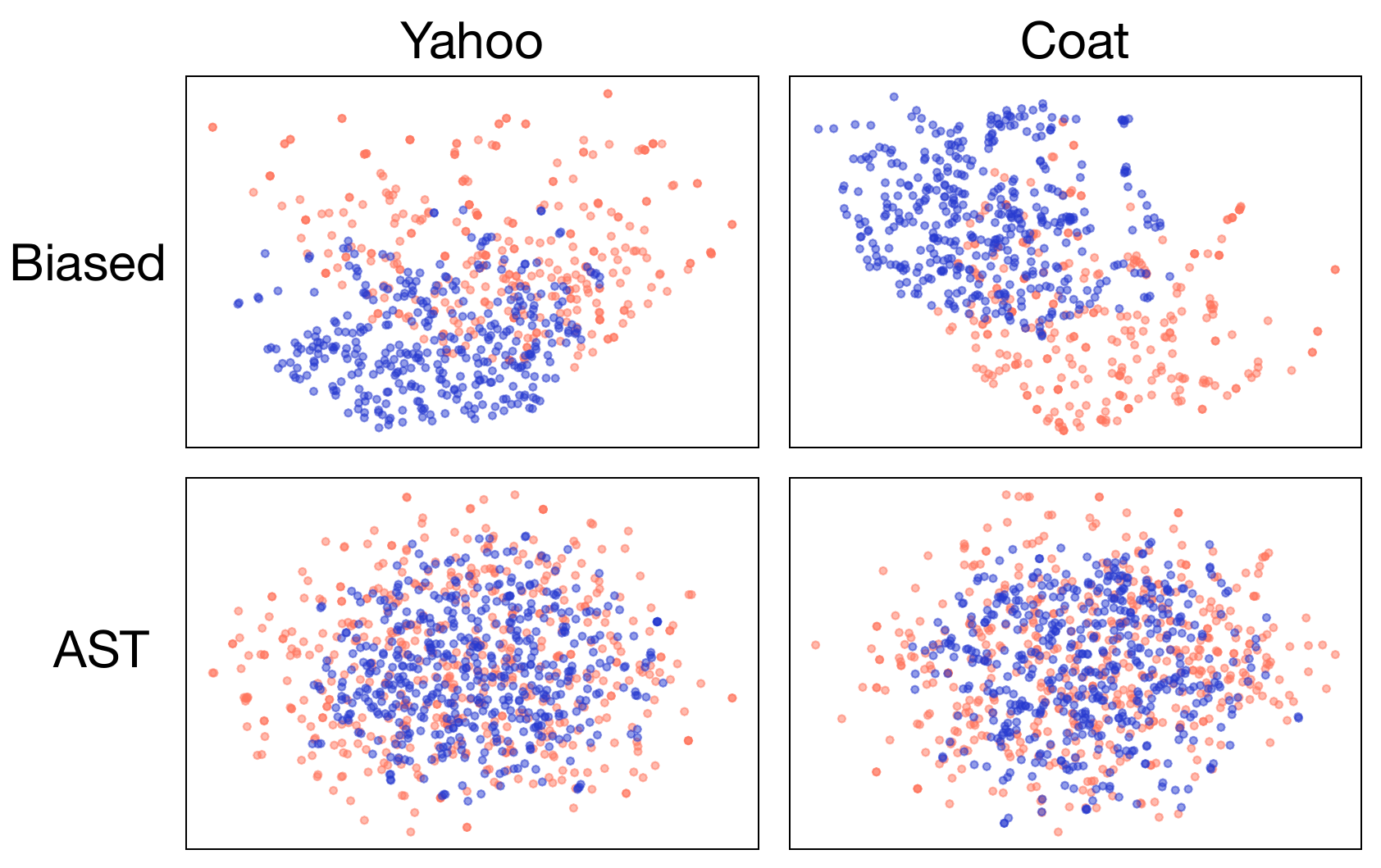}}
        \vskip -2em
     \caption{\small Visualization for t-SNE embeddings of AST’s and Biased’s representations. The blue points
correspond to the biased training data, while red ones correspond to  unbiased testing data.}\label{fig:TSNE}
    \vskip -1.5em
\end{figure}

\textbf{Setup}. We take a deeper examination on AST to understand what insights it can provide. We follow the same setup in \S~\ref{sec:overall}.\\
\textbf{Generalization and Convergence}. To understand the generalization and convergence of AST, Fig.~\ref{fig:loss} shows the curves of the training losses of different components and the testing NDCG on two datasets. We observe that: (\textbf{i}) AST is generally training-stable and can consistently converge and improve the unbiased testing performance with iterations. (\textbf{ii}) NDCG almost monotonically increases with iterations. This empirically suggests that minimizing our loss, i.e., the upper bound of the ideal loss is a valid approach to improve the ranking accuracy from biased feedbacks.\\
\textbf{Unobserved Confounders}. One of the most important properties of AST is to alleviate both marginal and conditional shifts caused by unobserved confounders. Thus, we investigate if the proposed AST has this ability and which component of AST plays the most important role. Fig.~\ref{fig:distancNCF} shows the empirically calculated $\mathcal{A}$-distance~\cite{long2018conditional} and MDD~\cite{li2020maximum} by using the learned embeddings of AST. Note that $\mathcal{A}$-distance and MDD can quantify covariate and concept shifts, respectively. From Fig.~\ref{fig:distancNCF}, we can find (\textbf{i}) $\mathcal{A}$-distance and MDD on AST embeddings are much smaller than those on vanilla NCF, suggesting that AST can reduce both covariate and concept distribution shifts more effectively. (\textbf{ii}) AST w/o D has a smaller $\mathcal{A}$-distance than  AST w/o A while AST w/o A has smaller MDD than  AST w/o D. This observation matches our idea that adversarial matching can minimize covariate shift while self-training can alleviate concept shift. To further intuitively understand the feature adaptation, we visualize the t-SNE embeddings sampled from  $P(\mathbf{z}_{ui})$ and  $Q(\mathbf{z}_{ui})$. From Fig.~\ref{fig:TSNE}, we can find AST effectively bridges the feature gap across biased data and unbiased data, but biased training fails since embeddings are separated and have a certain distance.

\section{Conclusions}

In this paper, we studied the problem of unbiased recommendation. We provided a novel perspective on the distribution shift for the unbiased recommendation problem. We derived several generalization bounds and presented both theoretical and algorithmic analyses of current learning algorithms. We also proposed the AST algorithm, which effectively addresses the issues of selection bias and unobserved confounders. Extensive experiments on three datasets with various settings demonstrated the effectiveness of AST. While our results strongly advocate for considering unobserved confounders in unbiased recommendation, optimizing them directly in the real world poses a challenge. Exploring more effective optimization methods is an interesting future research topic.


\clearpage
\bibliographystyle{ACM-Reference-Format}
\bibliography{Reference}
\onecolumn
\appendix

\section{The  Lemmas}
Before we conduct the proof, we first state the following Lemmas:
\begin{lemma}
\label{lemma_hh}
\cite{blitzer2007learning}. Let $\mathcal{H}$ be a hypothesis space of VC-dimension d, and for any distribution $P$ and $Q$  over $\mathcal{X}_{u} \times \mathcal{X}_{i}$, then $\forall h, h' \in \mathcal{H}$:
\begin{linenomath}
\begin{align}
|\mathcal{L}_{P}(h,h')-\mathcal{L}_{Q}(h,h')| \leq \frac{1}{2}d_{\mathcal{H} \Delta \mathcal{H}}(P, Q),
\end{align}
\end{linenomath}
where $d_{\mathcal{H} \Delta \mathcal{H}}(P, Q)=2 \sup _{h, h^{\prime} \in \mathcal{H}}|\mathbb{E}_{P}[\ell(h(\mathbf{x}_u,\mathbf{x}_i),h'(\mathbf{x}_u,\mathbf{x}_i))]-\mathbb{Q}[\ell(h(\mathbf{x}_u,\mathbf{x}_i),h'(\mathbf{x}_u,\mathbf{x}_i))]$. 
\end{lemma}

\begin{lemma}
\label{lemma_vapnik}
\cite{vapnik1999overview}. Let $S$ is a arbitrarily data distribution and $\mathcal{H}$ be a hypothesis space of VC-dimension d. Then $\forall h \in \mathcal{H}, \forall \delta>0$, w.p. at least $1-\delta$ over the  a sample size $N$ and natural exponential $e$:
\begin{linenomath}
\begin{align}
\mathcal{L}_{S}(h) \leq \widehat{\mathcal{L}}_{S}(h)+\sqrt{\frac{4}{N}(d \log \frac{2 e N}{d}+\log \frac{4}{\delta})}.
\end{align}
\end{linenomath}
\end{lemma}

\section{Proof of Theorem~\ref{theorem:multi-task}}
\label{Appendix:multi-task}

\begin{proof}
Following the definitions in \S~\ref{Pre}, we have:
\begin{linenomath}
\begin{align}
|\mathcal{L}_{P}(f)-\mathcal{L}_{Q}(f)|=|\mathcal{L}_{P}(f,k)-\mathcal{L}_{Q}(f,g)|,
\end{align}
\end{linenomath}
which has the following upper bound:
\begin{linenomath}
\begin{align}
|\mathcal{L}_{P}(f, k)-\mathcal{L}_{Q}(f, g)|&= |\mathcal{L}_{P}(f, k)-\mathcal{L}_{P}(f, g)+\mathcal{L}_{P}(f, g)-\mathcal{L}_{Q}(f, g)| \nonumber \\
&\leq  |\mathcal{L}_{P}(f, k)-\mathcal{L}_{P}(f, g)|+|\mathcal{L}_{P}(f, g)-\mathcal{L}_{Q}(f, g)| \nonumber \\
&=|\mathbb{E}_{P}[|f(\mathbf{x}_u,\mathbf{x}_i)-k(\mathbf{x}_u,\mathbf{x}_i)|-|f(\mathbf{x}_u,\mathbf{x}_i)-g(\mathbf{x}_u,\mathbf{x}_i)|]|  \label{EQ:1} \\
&+|\mathcal{L}_{P}(f, g)-\mathcal{L}_{Q}(f, g)|\leq \mathbb{E}_{P} [|k(\mathbf{x}_u,\mathbf{x}_i)-g(\mathbf{x}_u,\mathbf{x}_i)|]+\frac{1}{2}d_{\mathcal{H} \Delta \mathcal{H}}(P, Q),\nonumber 
\end{align}
\end{linenomath}
where we utilize the triangular inequality and
Lemma~\ref{lemma_hh}. Similarly, due to the symmetric property, the following inequality for  $Q$ holds:
\begin{linenomath}
\begin{align}
&|\mathcal{L}_{P}(f)-\mathcal{L}_{Q}(f)| =|\mathcal{L}_{P}(f, k)-\mathcal{L}_{Q}(f, g)| \leq \mathbb{E}_{Q} [|k(\mathbf{x}_u,\mathbf{x}_i)-g(\mathbf{x}_u,\mathbf{x}_i)|]+\frac{1}{2}d_{\mathcal{H} \Delta \mathcal{H}}(P, Q).\label{EQ:2}
\end{align}
\end{linenomath}
Combine the inequalities (\ref{EQ:1}) and (\ref{EQ:2}) above, we have:
\begin{linenomath}
\begin{align}
&\mathcal{L}_{Q}(f)\leq \mathcal{L}_{P}(f)+\frac{1}{2}d_{\mathcal{H} \Delta \mathcal{H}}(P, Q)+ \min \{\mathbb{E}_{P}[|k(\mathbf{x}_u,\mathbf{x}_i)-g(\mathbf{x}_u,\mathbf{x}_i)|], \mathbb{E}_{Q}[|k(\mathbf{x}_u,\mathbf{x}_i)-g(\mathbf{x}_u,\mathbf{x}_i)|]\}.  \label{ineq:1}
\end{align}
\end{linenomath}
Combining Eqs.~(\ref{EQ:2}), (\ref{ineq:1}) and Lemma~\ref{lemma_vapnik}, and considering the hypothesis $f(\mathbf{x}_{u},\mathbf{x}_{i})$ is composed of a two parts: $f=h \circ \phi$ where $h$ is the hypothesis and $\phi$ maps $(\mathbf{x}_{u},\mathbf{x}_{i})$ to $\mathbf{z}_{ui}$. W.p. at least $1-\delta$:
\begin{linenomath}
\begin{align}
&\mathcal{L}_{Q}(h \circ \phi) \leq  \mathcal{\widehat{L}}_{P}(h \circ \phi)+\frac{1}{2} d_{\mathcal{H} \Delta \mathcal{H}}({{P}}(\mathbf{z}_{ui}), {{Q}}(\mathbf{z}_{ui})) \label{Eq:appendixmulti} \\
&+\min \{\mathbb{E}_{P(\mathbf{z}_{ui})}[|\tilde{k}(\mathbf{z}_{u})-\tilde{g}(\mathbf{z}_{ui})|], \mathbb{E}_{Q({\mathbf{z}_{ui})}}[|\tilde{k}(\mathbf{z}_{ui})-\tilde{g}(\mathbf{z}_{ui})|]\} +\sqrt{\frac{4}{N}\big(d \log \frac{2 e N}{d}+\log \frac{4}{\delta}\big)} \nonumber,
\end{align}
\end{linenomath}
where ${{P}}(\mathbf{z}_{ui})$ (resp. ${{Q}}(\mathbf{z}_{ui}))$ is the probability density functions over $\mathcal{Z}$ induced by $P(\mathbf{x}_u, \mathbf{x}_{i},O_{ui})$ (resp. $Q(\mathbf{x}_{u},\mathbf{x}_{i},O_{ui})$) and $\phi$. $\tilde{g}={\int_{\phi^{-1}_{(z)}} g(x) p(x) d x}/{\int_{\phi^{-1}_{(z)}} p(x) d x}$ where $x$ denotes the features \cite{johansson2019support} (resp. $\tilde{k}$) is the latent labeling function induced by $g$ (resp. $k$) and $\phi$: With Lemma~\ref{lemma_vapnik} and $f=h \circ \phi$, we have, w.p. at least $1-\delta$:
\begin{linenomath}
\begin{align}
\mathcal{L}_{Q}(h \circ \phi) \leq \widehat{\mathcal{L}}_{Q}(h \circ \phi)+\sqrt{\frac{4}{N}(d \log \frac{2 e N}{d}+\log \frac{4}{\delta})}.\label{Eq:appendixQ}
\end{align}
\end{linenomath}
We can completes the whole proof by combining Eq.~\eqref{Eq:appendixQ} with Eq.~\eqref{Eq:appendixmulti} over coefficients $\rho$ and $1-\rho$, respectively.
\end{proof}

\section{Proof of Theorem~\ref{theorem:meta}}
\label{Appendix:meta}
\begin{proof}
We firstly  derive the upper bound between the expected unbiased error $\mathcal{L}_{Q}(h)$ and the empirical unbiased error $\widehat{\mathcal{L}}_{Q}(h(w))$ via the meta validation. Specifically, we define that $\epsilon_{i}(h(w))=\mathcal{L}_{Q}(h)-\ell (h(w)(\mathbf{x}_{u},x_i), Y_{ui})$ for $h(w) \in \mathcal{H}^{\prime}$ and every data sample in $(\mathbf{x}_u,\mathbf{x}_i,Y_{ui})\in \mathcal{D}_{Q}$. Then, we have:
\begin{linenomath}
\begin{align}
\mathcal{L}_{Q}(h)-\mathcal{\widehat{L}}_{Q}(h(w))=\frac{1}{M} \sum_{m=1}^{M} \epsilon_{m}(h(w)).
\end{align}
\end{linenomath}
Since  $\mathcal{L}_{Q}(h) \in[0,1]$ and $\ell(h(w)(\mathbf{x}_{u},x_i), y) \in [0,1]$, we have $\mathcal{L}_{Q}(h)-\ell(h(w)(\mathbf{x}_{u},x_i), y) \in[-1,1]$, $\mathbb{E}[\epsilon_{m}(h(w))^{2}] \leq 1$, and $|\epsilon_{m}(h(w))| \leq 1$. Based on the Bernstein inequality~\cite{uspensky1937introduction}, we have:
\begin{linenomath}
\begin{align}
p(\frac{1}{M} \sum_{m=1}^{M} \epsilon_{m}(h(w))>\xi) \leq \exp (-\frac{\xi^{2} M / 2}{1+\xi / 3}). \label{union}
\end{align}
\end{linenomath}
Taking the union bound of this inequality over all $h(w) \in \mathcal{H}^{\prime}$ has:
\begin{linenomath}
\begin{align}
p(\cup_{h(w) \in \mathcal{H}^{\prime}} \frac{1}{M} \sum_{m=1}^{M} \epsilon_{m}(h(w))>\xi) \leq M^{d^{\prime}} \exp (-\frac{\xi^{2} M / 2}{1+\xi / 3}). \label{delta}
\end{align}
\end{linenomath}
Let $\delta=M^{d^{\prime}} \exp (-\frac{\xi^{2} M / 2}{1+\xi / 3})$. Solving the above Inequality (\ref{delta}) for $\xi$ yields the following result (note that $\xi \geq 0$):
\begin{linenomath}
\begin{align}
&\xi=\frac{d^{\prime} \log M-\log \delta}{3M} \pm \sqrt{(\frac{d^{\prime} \log M-\log \delta}{3M})^{2}+\frac{2(d^{\prime} \log M-\log \delta)}{M}}\leq  \frac{d^{\prime} \log M-\log \delta}{3 M}+\sqrt{\frac{2(d^{\prime} \log M-\log \delta)}{M}} \because \sqrt{a+b} \leq \sqrt{a}+\sqrt{b}.
\end{align}
\end{linenomath}
Thus, for any $\delta>0$, with probability at least $1-\delta$, for $h' \in \mathcal{H}^{\prime}$,
\begin{linenomath}
\begin{align}
\mathcal{L}_{Q}(h) \leq \mathcal{\widehat{L}}_{Q}(h(w))+ \frac{d^{\prime} \log M-\log \delta}{3M}+\sqrt{\frac{2(d^{\prime} \log M-\log \delta)}{M}}.
\end{align}
\end{linenomath}
Similar to Eq.~\eqref{Eq:appendixQ}, by furthering considering the above bound in the latent feature space via $\phi$ and combine it with  Eq.~\eqref{Eq:appendixmulti} over coefficients $\rho$ and $1-\rho$ respectively, we complete the proof.
\end{proof}
\section{Proof of Theorem~\ref{theorem:KL}}
\label{appendix:TheoremKL}
\begin{proof}
We show that the ideal risk $\mathcal{L}_{Q}(f)=\mathcal{L}_{Q_z}(h)$ can bounded as (note that we denote $P(\mathbf{z}_{ui})$ $(Q(\mathbf{z}_{ui}))$  as $P_z$ ($Q_z$) for brevity):
\begin{linenomath}
\begin{align}
\mathcal{L}_{Q}(f)&=\mathcal{L}_{Q_z}(h)=\mathcal{L}_{Q_z}(h)-\mathcal{L}_{P_z}(h,\tilde{k})+\mathcal{L}_{P_z}(h,\tilde{k})-\mathcal{L}_{P_z}(h)+\mathcal{L}_{P_z}(h)\nonumber\\
&\leq \mathcal{L}_{P_z}(h) +|\mathcal{L}_{P_z}(h)-\mathcal{L}_{P_z}(h,\tilde{k})|+|\mathcal{L}_{Q_z}(h)-\mathcal{L}_{P_z}(h,\tilde{k})| \nonumber\\
&=\mathcal{L}_{P_z}(h)+|\mathbb{E}_{P_z}[|h(z)-\tilde{g}(z)|-|h(z)-\tilde{k}(z)|]|+|\mathcal{L}_{Q_z}(h)-\mathcal{L}_{P_z}(h,\tilde{k})|\nonumber\\
&\leq \mathcal{L}_{P_z}(h)+\mathbb{E}_{P_z}[|\tilde{g}(z)-\tilde{k}(z)|]+|\mathcal{L}_{Q_z}(h)-\mathcal{L}_{P_z}(h,\tilde{k})|  \nonumber\\
& \leq \mathcal{L}_{P_z}(h)+\mathbb{E}_{P_z}[|h(z)-\tilde{k}(z)|]+\int |P_z-Q_z|\cdot|h(z)-\tilde{k}(z)|dz \nonumber  \\
& \leq \mathcal{L}_{P_z}(h)+\mathbb{E}_{P_z}[|\tilde{g}(z)-\tilde{k}(z)|]+\int |P_z-Q_z|dz \because h(z)-\tilde{k}(z) \in [0,1]\nonumber\\
&=\mathcal{L}_{P_z}(h)+\mathbb{E}_{P_z}[|\tilde{g}(z)-\tilde{k}(z)|]+{\mathrm{TV}}(P_z||Q_z). \nonumber \\
&\leq \mathcal{L}_{P_z}(h)+\mathbb{E}_{P_z}[|\tilde{g}(z)-\tilde{k}(z)|]+\sqrt{2{\mathrm{KL}}({{P}}_{z} \| {{Q}}_{z})}, \label{Eq:5.1-1}
\end{align}
\end{linenomath}
where we used triangular inequality multi-times and the Pinsker’s inequality~\cite{csiszar2011information} in the last line. $h(z)-\tilde{k}(z) \in [0,1]$ since our loss is 0-1 binary loss. Due to the the symmetric property, we also have:
\begin{align}
\mathcal{L}_{Q_z}(h) \leq \mathcal{L}_{P_{z}}(h)+\mathbb{E}_{Q}[|\tilde{g}(z)-\tilde{k}(z)|]+\sqrt{2 {\mathrm{KL}}({{P}}_{z} \|{{Q}}_{z}}), \label{Eq:5.1-2}
\end{align}
Combining Eqs.~(\ref{Eq:5.1-1}), (\ref{Eq:5.1-2}) and Lemmas~\ref{lemma_vapnik}, we have:
\begin{linenomath}
\begin{align}
&\mathcal{{L}}_{Q}(h  \circ \phi) \leq  \mathcal{\widehat{L}}_{P}(h \circ \phi ) +\frac{1}{2}  \sqrt{2 {\mathrm{KL}}({{P}}(\mathbf{z}_{ui}) \| {{Q}}(\mathbf{z}_{ui}))}+\rho\min \{\mathbb{E}_{P(\mathbf{z}_{ui})}[|\tilde{g}(\mathbf{z}_{ui})-\tilde{k}(\mathbf{z}_{ui})|], \mathbb{E}_{Q(\mathbf{z}_{ui})}[|\tilde{g}(\mathbf{z}_{ui})-\tilde{k}(\mathbf{z}_{ui})|]\}\nonumber
\end{align} 
\end{linenomath}
We can complete the proof by summing this bound with Eq.~\eqref{Eq:appendixQ} over coefficients $\rho$ and $1-\rho$, respectively.
\end{proof}
\section{The Experimental  Settings}
\label{appendix:exp}
\textbf{Implementations}. For most of baselines, we use the official implementations publicly released by the authors on Github:\\
\textbf{IPS}: \url{https://github.com/usaito/unbiased-implicit-rec}\\
\textbf{CVIB}: \url{https://github.com/RyanWangZf/CVIB-Rec}\\
\textbf{KD}: \url{https://github.com/dgliu/SIGIR20_KDCRec}\\
 \textbf{ACL}: \url{https://github.com/StatsDLMathsRecomSys/Adversarial-Counterfactual-Learning-and-Evaluation-for-Recommender-System} \\
 \textbf{ATT}: \url{https://github.com/usaito/asymmetric-tri-rec-real}\\
\textbf{AutoDebis}: \url{https://github.com/DongHande/AutoDebias}

\end{document}